\documentclass[a4paper,fleqn,usenatbib]{mnras}

\usepackage{newtxtext,newtxmath}
\usepackage[T1]{fontenc}
\usepackage{ae,aecompl}

\usepackage{graphicx}	% Including figure files
\usepackage{amsmath}	% Advanced maths commands
\usepackage{amssymb}	% Extra maths symbols
\usepackage{amsfonts}
\title[Weak Lensing Analysis of CODEX clusters]
{Weak Lensing Analysis of CODEX Clusters using Dark Energy Camera Legacy Survey : Mass-Richness Relation}

\author[A. Phriksee et al.]
{Anirut Phriksee$^{1, 2}$,\thanks{E-mail: anirut\_phriksee@cmu.ac.th}
Eric Jullo$^{1}$,
Marceau Limousin$^{1}$,
HuanYuan Shan$^{3}$,
\newauthor
Alexis Finoguenov$^{4}$,
Siramas Komonjinda$^{5, 6}$,
Suwicha Wannawichian$^{5, 6}$,
\newauthor
Utane Sawangwit$^{7}$
\\
$^{1}$Aix Marseille Univ, CNRS, CNES, LAM, Marseille, France\\
$^{2}$Ph.D. Program in Physics, Department of Physics and Materials Science, Faculty of Science, 
Chiang Mai University, Chiang Mai, Thailand\\
$^{3}$Shanghai Astronomical Observatory, Shanghai, China\\
$^{4}$Department of Physics, University of Helsinki, Helsinki, Finland\\
$^{5}$Department of Physics and Materials Science, Faculty of Science, Chiang Mai University, 
Chiang Mai, Thailand\\
$^{6}$Research Center in Physics and Astronomy, Faculty of Science, Chiang Mai University, Chiang Mai, Thailand\\
$^{7}$National Astronomical Research Institute of Thailand, Chiang Mai, Thailand
}

\date{Accepted XXX. Received YYY; in original form ZZZ}

\pubyear{2019}

\begin{document}
\label{firstpage}
\pagerange{\pageref{firstpage}--\pageref{lastpage}}
\maketitle

% Abstract of the paper
\begin{abstract}
We present the weak lensing analysis of 279 CODEX clusters 
using imaging data from 4200 $\text{deg}^{2}$ of the DECam Legacy Survey (DECaLS) Data Release 3.
The cluster sample results from a joint selection in X-ray, optical richness in the range 20 $\leq$ $\lambda$ < 110, and redshift in the range 0.1 $\leq$ $z$ $\leq$ 0.2.
We model the cluster mass ($M_{\rm 200c}$) and the richness relation with the expression 
$\left\langle M_{\rm 200c} | \lambda \right\rangle \propto M_{0} \, (\lambda / 40)^{F_{\lambda}}$. 
By measuring the CODEX cluster sample as an individual cluster, we obtain the best-fit values,
$M_{0} =  3.24^{+0.29}_{-0.27} \times 10^{14} \text{M}_{\odot}$, and 
 $F_{\lambda} = 1.00 ^{+0.22}_{-0.22}$ for the richness scaling index, consistent with a power law relation.
Moreover, we separate the cluster sample into three richness groups; $\lambda = $ 20 - 30, 30 - 50 and 50 - 110, and measure the stacked excess surface mass density profile in each group.
The results show that both methods are consistent.
In addition, we find an excellent agreement between our weak lensing based scaling relation and the relation obtained with dynamical masses estimated from cluster member velocity dispersions measured by the SDSS-IV/SPIDERS team. 
This suggests that the cluster dynamical equilibrium assumption involved in the dynamical mass estimates is statistically robust for a large sample of clusters.
\end{abstract}

% Select between one and six entries from the list of approved keywords.
% Don't make up new ones.

%%%%%%%%%%%%%%%%%%%%%%%%%%%%%%%%%%%%%%%%%%%%%%%%%%

%%%%%%%%%%%%%%%%% BODY OF PAPER %%%%%%%%%%%%%%%%%%

%%%%%%%%%%%%%%%%%%%%%%%%%%%%%%%%%%%%%%%%%%%%%%%%%
%%%%%%%%    Section I : Introduction    %%%%%%%%%
%%%%%%%%%%%%%%%%%%%%%%%%%%%%%%%%%%%%%%%%%%%%%%%%%
\section{Introduction}\label{sec:introduction}
Presently, the most accurate model to describe the universe is called 
the Lambda-Cold Dark Matter ($\Lambda$CDM) model, 
also know as the standard cosmological model.
Observations suggest that baryons represent a few percent 
of the energy content of matter in the universe. 
The rest, identified as dark energy and dark matter, 
remains today relatively unexplained \citep[e.g.][]{planck2016}.
Dark energy is responsible for the recent acceleration in the expansion of the universe. 
Assuming it would be homogeneous and isotropic,
the accelerated expansion would imply this medium to have a negative pressure. 
Other models rather propose to modify the equations of general relativity 
on the largest scales of the universe to account for the accelerating expansion.
Dark matter, probably an elementary particle still to be discovered, 
accumulates in large structures of the universe, such as galaxies, 
clusters of galaxies and cosmic filaments under the effect of gravity.

Galaxy clusters, the last structures formed in the universe, 
are very exciting objects for cosmological studies. 
First, they have a well-defined mass function, 
whose shape reflects the value of cosmological parameters (\citealp{press1974, warren2006, tinker2008, bhattacharya2011, despali2016, mcclintock2019apj}).
Second, numerical simulations predict characteristic density profile (e.g., NFW profile; \citealp{navarro1996, nfw1997, limousin2013, diemer2014, bose2017}, Einasto profile; \citealp{einasto1965, hayashi2008, diemer2014}), 
sub-halo mass function and triaxiality (\citealp{becker2011, angulo2012, noh2012}), which depend on the properties of the dark matter particles, although in reality, astrophysical effects tend to mask these effects \citep[e.g.][]{newman2013}.
Finally, their gas-mass ratios reflect the cosmological content \citep[e.g.][]{allen2008, mant2016}.

Due to their complex internal physical processes, and their extreme masses, galaxy-clusters have been observed at all wavelengths from X-ray to radio, and large surveys now give us a comprehensive view of their distribution and evolution with time. Cluster mass can be determined from their X-ray brightness, temperature (\citealp[e.g.][]{sarazin1986, rosati2002, donahue2014}), 
or Sunyaev-Zel'dovich effect (SZE) in the radio domain
(\citealp[e.g.][]{baxter2015, planck2016SZE, geach2017, raghunathan2019}).

Galaxy-cluster richness provides another estimate for cluster masses, based on the number of cluster-member galaxies. 
The red-sequence Matched-filter Probabilistic Percolation (redMaPPer) method determines 
cluster-richness from the number of galaxies in an aperture, distributed in a red-sequence in a color-magnitude diagram \citep{rykoff2014}.
The tight correlation between cluster richness and masses determined by other methods 
demonstrates the reliability of this estimator (\citealp{saro2015, saro2017, farahi2016, mantz2016, simet2017, melchior2017, murata2018, murata2019, mcclintock2019}).

Assuming an NFW mass profile and a range of anisotropy profiles, it is also possible to determine cluster masses with the Jean equation and spectroscopic redshifts of cluster-member galaxy 
(\citealp{biviano2013, mamon2013, munari2013}).
This estimation is based on the assumption that galaxy clusters are in the hydrostatic equilibrium state.
However, some biases might affect the estimated mass due to complex physical processes or the non-equilibrium states in galaxy clusters, for instance, the infall motion of outer galaxy members \citep{falco2013}.
The cross study between the dynamical analysis and weak lensing analysis helps to probe the assumption that galaxy clusters are in the equilibrium state \citep[e.g.][]{smith2016}.

Among all the methods to estimate cluster masses, gravitational lensing is recognized 
as the one with the least amount of assumptions regarding physical processes. 
In this respect, many methods are calibrated on gravitational lensing measurements 
(\citealp[e.g.][]{johnston2007, mandelbaum2006}). Recently, the mass-richness relation of redMaPPer clusters has 
been estimated in the 10,000 deg$^2$ of the Sloan Digital Sky Survey (\citealp{zu2015, simet2017, murata2018}), 
the Dark Energy Survey Science Verification (DES-SV) data \citep{melchior2017} 
and with the Dark Energy Survey Year 1 data  
\citep{mcclintock2019}.

Weak-lensing analysis requires high quality and deep imaging data.
The measured galaxy ellipticities are affected by different kinds of noises (telescope, instrument, atmosphere, etc.) that need to be taken into account. Building on the experience of previous experiments
(\citealp[e.g.][]{mandelbaum2005, heymans2012, miller2013}),
we now have several procedures to characterize and correct these effects.
Weak lensing analysis also requires a good knowledge of source redshift distribution.
Photometric redshifts require multi-band imaging, but also calibration techniques, 
for instance, based on spectroscopic surveys
(e.g. \cite{laigle2016} in the COSMOS field).

In this paper, we present a weak lensing analysis of the COnstrain Dark Energy
with X-ray galaxy clusters (CODEX) sample. 
We produce our lensing catalog, based on the photometric catalogs produced by 
the Dark Energy Camera Legacy Survey (DECaLS) Data Release 3. This imaging survey
covers 4200 deg$^2$ in $grz$ bands, and partly overlaps with the CODEX catalog footprint.
The CODEX cluster sample is the intersection between the redMaPPer cluster sample, and 
clusters identified by detecting overdensities in X-ray observations
from the ROSAT All-Sky Survey \citep[RASS,][]{voges1999}. This catalog is supposed to be 
less contaminated by fake detections, and thus more reliable to constrain cosmological parameters.
We measure the lensing signal around 279 CODEX clusters selected in the redshift range $0.1 \leq z \leq 0.2$. 
We use these measurements to estimate the parameters of the scaling relation 
that relates to the lensing mass and the richness of the clusters.
We compare our scaling relation based on weak lensing with the one determined from the dynamics of 
the galaxies observed in spectroscopy presented in \citet{capasso2019}.

The structure of this paper is as follows.
In Section \ref{sec:data}, we describe the preparation of the DECaLS data for the shear measurement
and the CODEX cluster catalog.
We summarize the method used for the weak lensing analysis in Section \ref{sec:wl} and the data analysis in Section \ref{sec:analysis}.
Section \ref{sec:results} reports on the results from the weak lensing measurement 
and also the comparison with the dynamical analysis in Section \ref{sec:wl-dynamic}.
In Section \ref{sec:conclusion}, we conclude and discuss future work and analysis.

Throughout the paper, we adopt a concordance $\Lambda$CDM model with
the cosmological values :
$\Omega_{m}$ = 0.3,
$\Omega_{\Lambda}$ = 0.7,
$\sigma_{8}$ = 0.8,
$n_{s}$ = 0.965
and a Hubble constant $H_{0}$ = 100 $h$ km s$^{-1}$ Mpc$^{-1}$ with $h$ = 0.7. Errors are quoted on the 68\% confidence level.

%%%%%%%%%%%%%%%%%%%%%%%%%%%%%%%%%%%%%%%%%%%%%%%%%
%%%%%%%%%%      Section II : DATA    %%%%%%%%%%%%
%%%%%%%%%%%%%%%%%%%%%%%%%%%%%%%%%%%%%%%%%%%%%%%%%
\section{DATA}\label{sec:data}
In this study, we use the data from the Dark Energy Camera Legacy Survey (DECaLS) Data Release 3 
overlapping with the CODEX cluster catalog.
We describe the preparation of the DECaLS shear catalog from the DECaLS DR3 data in Section \ref{sec:decals}, and 
the CODEX clusters catalog in Section \ref{sec:codex}.
\begin{figure}
\includegraphics[width=0.98\columnwidth]{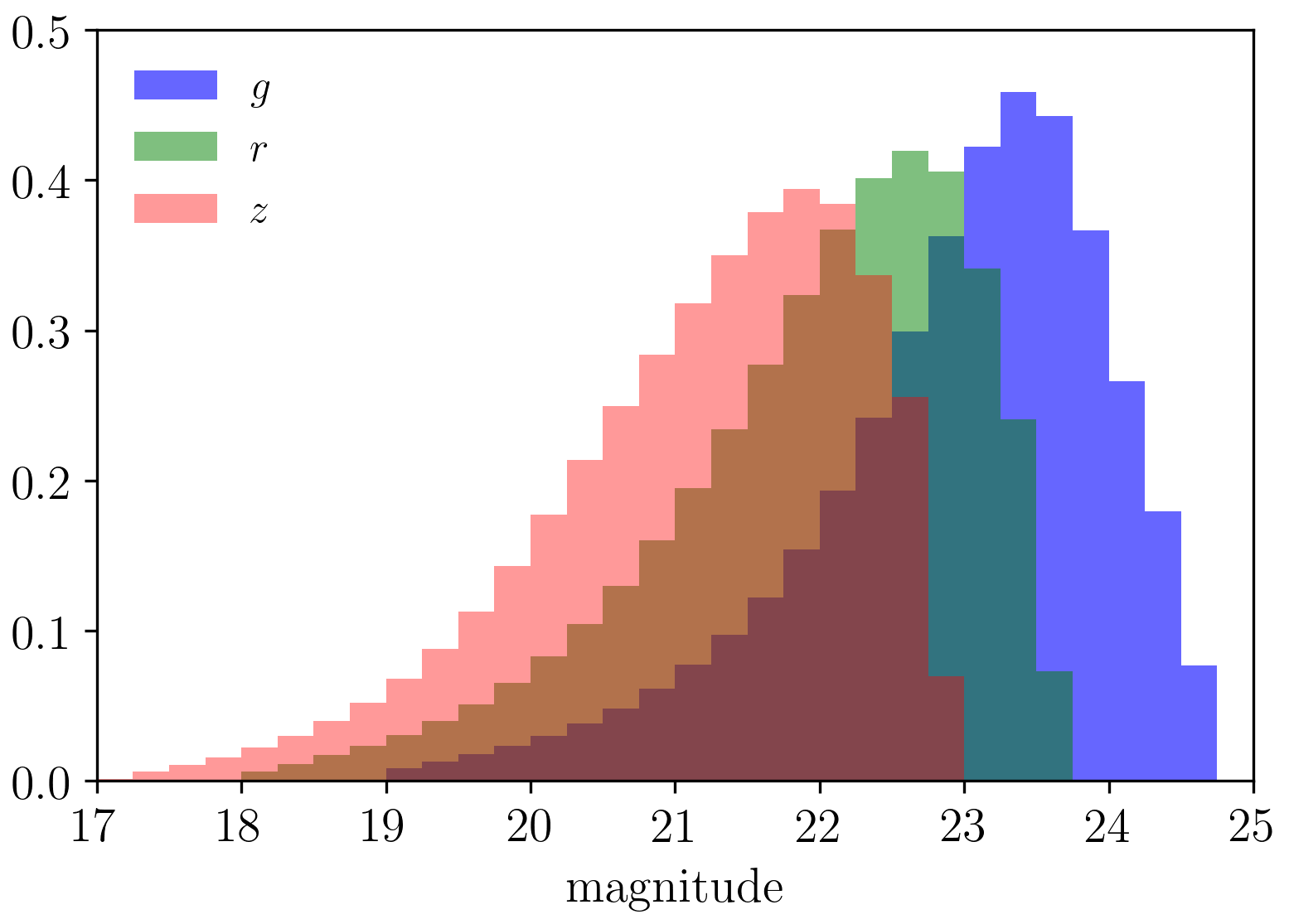}
\caption{The depth of the DECaLS DR3 objects in three optical bands from the DECaLS DR3.}
\label{fig:magnitude_grz}
\end{figure}
\begin{figure*}
	\includegraphics[width=\textwidth]{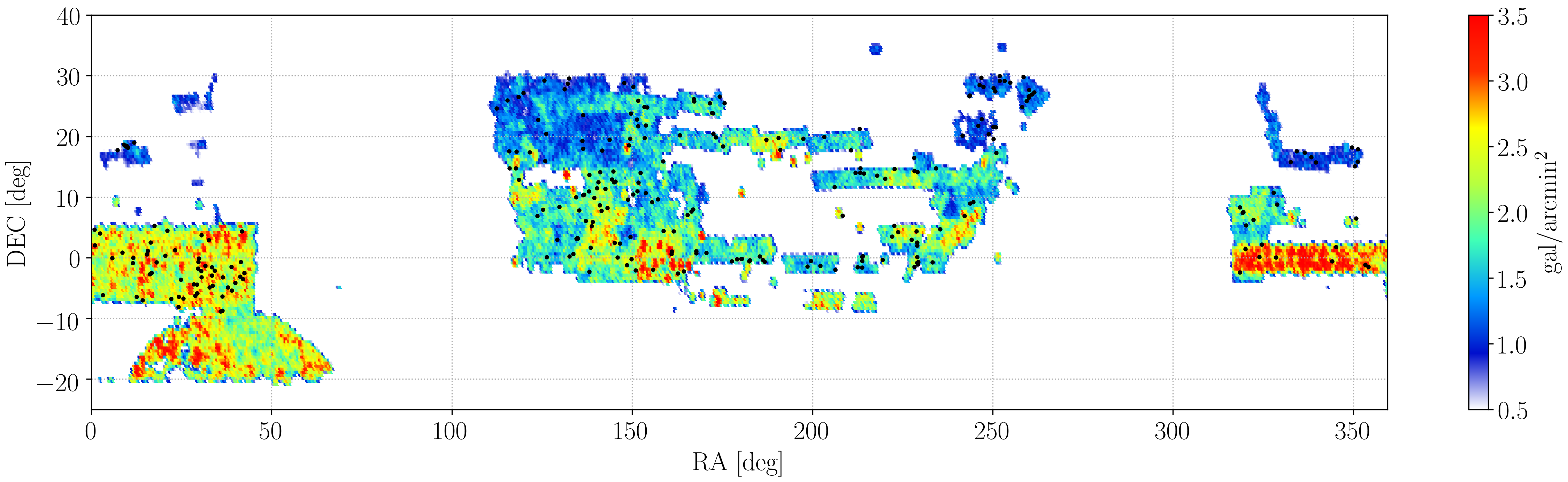}
    \caption{A source density distribution of the DECaLS DR3 shear catalog and the samples of CODEX clusters (black dots) in Equatorial coordinates.}
    \label{fig:sky_coverage}
\end{figure*}
\subsection{DECaLS}\label{sec:decals}
The Dark Energy Camera Legacy Survey (DECaLS)\footnote{\url{http://www.legacysurvey.org}} 
uses the Dark Energy camera installed on the Blanco 4 meters telescope, 
located at the Cerro Tololo Inter-American Observatory in Chile. 
This imaging survey is part of the Dark Energy Spectroscopic Instrument (DESI)
Legacy Imaging Survey \citep{dey2019}.
It provides the optical imaging for targeting 2/3 of the DESI footprint at declination DEC < 32. 
DR3 (PI: D. Schlegel and A. Dey)\footnote{DECam programs ID: 2013A-0741 and 2014B-0404} includes the images observed from August 2014 to March 2016 
and also incorporates data from the Dark Energy Survey (DES) 
acquired between September 2012 and March 2016.
We note that only DECaLS data were used for this study.
The DECaLS DR3 data contain the images covering 
4300 deg$^2$ in $g$-band, 
4600 deg$^2$ in $r$-band 
and 8100 deg$^2$ in $z$-band.
In total 4200 deg$^2$ has been observed in all three optical bands.

In Figure \ref{fig:magnitude_grz}, we show the depth of the DECaLS DR3 objects in three optical bands; 
$g$, $r$ and $z$. The DECaLS is about 1.5 to 2.0 magnitudes deeper than SDSS in $r$-band \citep{dey2019}.
Forced photometry is performed with the tool Tractor \citep{lang2014}.

In the DECaLS DR3 catalog, the sources from the Tractor catalog 
are separated into five morphological models;
Point sources (PSF), Simple galaxies (SIMP), DeVaucouleurs (DEV), Exponential (EXP), and Composite model (COMP).
Sky-subtracted images are stacked in five different ways: one stack per band, one ``flat" Spectral Energy Distribution (SED) stack of all bands, one ``red" SED stack of all bands ($g-r=1$ mag and $r-z=1$ mag). Sources above 6$\sigma$ detection limit in any stack are kept as candidates. PSF (delta function) and SIMP models are adjusted on individual images convolved by their own PSF models. PSF models for the individual exposure are determined with the tool PSFEx \citep{bertin2011}.

A source is retained if its penalized $\chi^2$ is improved by 25;
sources below this threshold are removed. Then, the source is classified
as the better of PSF or SIMP, unless adjusting a DEV or EXP profile improves the $\chi^2$ by 9 (approximately $3\sigma$ improvement).
A source is upgraded as COMP (composite between DEV and EXP model),
if the penalized $\chi^2$ improves by another 9. 
These selections imply that any extended source classification 
corresponds to at least a 5.8$\sigma$ detection, and a 6.5$\sigma$ detection for the COMP model.

\subsubsection{Bias correction}
Galaxy ellipticity parameters $\varepsilon_{1}$ and $\varepsilon_{2}$ are free parameters of the SIMP, DEV, EXP and COMP models.
They are estimated by a joint fit on the three optical $grz$ bands. 
We model potential measurement bias with a multiplicative ($m$) and additive bias ($c$) \citep[e.g.][]{heymans2012,miller2013}.
The additive bias is known to arise from residuals in the anisotropic PSF correction, and depends on galaxy sizes.
The multiplicative bias arises from shear measurement, 
which can be generated by many effects, such as measurement method, blending, and crowding.
\citep[see e.g.][]{martinet2019}. 
In order to calibrate the DECaLS DR3 shear catalog, 
we cross-match the DECaLS objects with the Canada France Hawaii Telescope (CFHT) Stripe 82 objects and compute the correction parameters.
The CFHT Stripe 82 (CS82) is a survey covering $\approx$ 170 square degrees of the Sloan Digital Sky Survey (SDSS) Stripe 82 in the equatorial region of the South Galactic Cap. Imaging data are of high quality and have been obtained in excellent seeing conditions between 0.4 - 0.8 arcsec with an average of 0.59 arcsec. It was primarily designed for lensing analysis (\citealp[see e.g.][]{huanyuan2014, liu2015}).

For each model SIMP, EXP and DEV, we compute the correction parameters defined by
\begin{eqnarray}
\varepsilon_{i}^\text{obs} = (1 + m_{i}) \, \varepsilon_{i}^\text{true} + c_{i} \,; i = 1, 2
\end{eqnarray}
where $\varepsilon^\text{obs}$ is the observed shape of the source
and $\varepsilon^\text{true}$ is the true shape.
For the multiplicative bias, we define the correction factor which is given by
\begin{eqnarray}\label{eq:oneplusm}
	1 + m = \frac{a_{0} \, \text{exp}(-a_{1} \times r_{g} \times magz)}
	{\text{log}_{10} magz},
\end{eqnarray}
where $r_{g}$ is a radius of the objects, $magz$ is a magnitude of the object in the $z$ band,
$a_{0}$ and $a_{1}$ are the result from the fitting with the CS82 data
described in Appendix \ref{sec:correction_factor}.
In Figure \ref{fig:sky_coverage}, we plot the sky coverage of the DECaLS DR3 shear catalog with the galaxy density.
We also compute the effective weighted galaxy number density \citep{heymans2012} 
of DECaLS galaxies which is similar to the distribution 
in Figure \ref{fig:sky_coverage},
higher density in South Galactic Cap (e.g., Stripe 82 region).
This is a little larger than the data from the Sloan Digital Sky Survey (SDSS) 
which has a source density of about 1.5 galaxies per square arcminute \citep{mandelbaum2005}.
We mention that many sources are discarded 
because Tractor is not specific to shape measurement.
In addition, we excluded the SIMP object in the DECaLS shear catalog for our weak lensing analysis because the result in calibrating with the CS82 provided high correction values, as shown in Table \ref{tab:ellipticity_decals}.
\begin{figure}
	\includegraphics[width=\columnwidth]{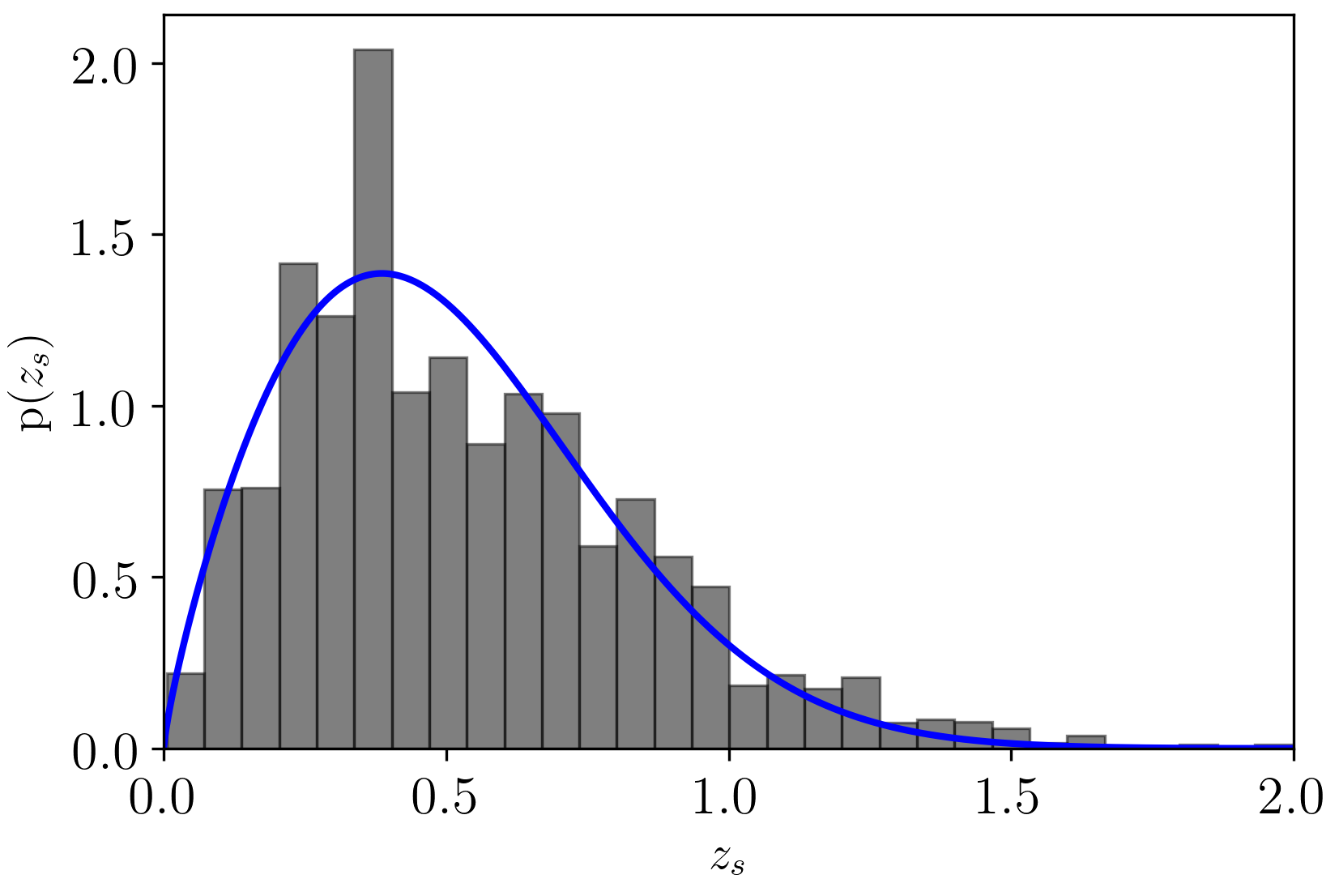}
    \caption{The photometric redshift distribution of DECaLS sources from the COSMOS 2015 catalog 
    with the probability function (blue solid line) defined by Equation \ref{eq:prob_zs}.}
    \label{fig:prob_pz}
\end{figure}

\subsubsection{Redshift distribution}
In this section, we describe how we compute the redshift distribution of the DECaLS sources. 
We use the photometric redshifts measured in the COSMOS field \citep{laigle2016}.
The photometric redshifts of the COSMOS 2015 catalog are computed by using the ``Le Phare" program with a chi-square fitting method between the theoretical and observed photometric catalog same as \citet{ilbert2013}.
We match each source of the DECaLS catalog in position with the COSMOS catalog, with maximum separation of 1 arcsec. 
We manage to match 100\% of the DECaLS shear catalog with the COSMOS 2015 catalog and compute the histogram of the photometric redshift distribution for the DECaLS DR3 shear catalog as shown in Figure \ref{fig:prob_pz}.
We model the photometric redshift distribution by fitting with the following probability distribution function 
\begin{eqnarray}\label{eq:prob_zs}
P(z_{s}) \propto A \, \left ( \frac{z_{s}}{z_0} \right ) ^{B - 1} \text{exp} \left[ -\frac{1}{2} \left( \frac{z_{s}}{z_0} \right) ^ {2} \right ] \, ,
\end{eqnarray}
and find good agreement as seen \ref{fig:prob_pz}.
We obtain the fitting parameters, 
$A = 2.261 \pm 0.172$, 
$B = 1.801 \pm 0.173$ 
and $z_{0} = 0.432 \pm 0.035$.
This function is used to calculate the excess mass density profile 
as described in Section \ref{sec:nfw_profile}.
In Figure \ref{fig:prob_pz}, we show the comparison
between the redshift distribution of the DECaLS sources from the COSMOS 2015 catalog and the probability function defined in Equation \ref{eq:prob_zs}.

In our analysis,  we consider all galaxies along the line-of-sight. Given the limited color information in our galaxy sample, we do not remove the foreground galaxies from the shear measurement. We did not find any selection in color to yield a higher signal-to-noise in the lensing measurement than the case of no removal of the foreground galaxies. 

\begin{figure}\centering
	\includegraphics[width=0.95\columnwidth]{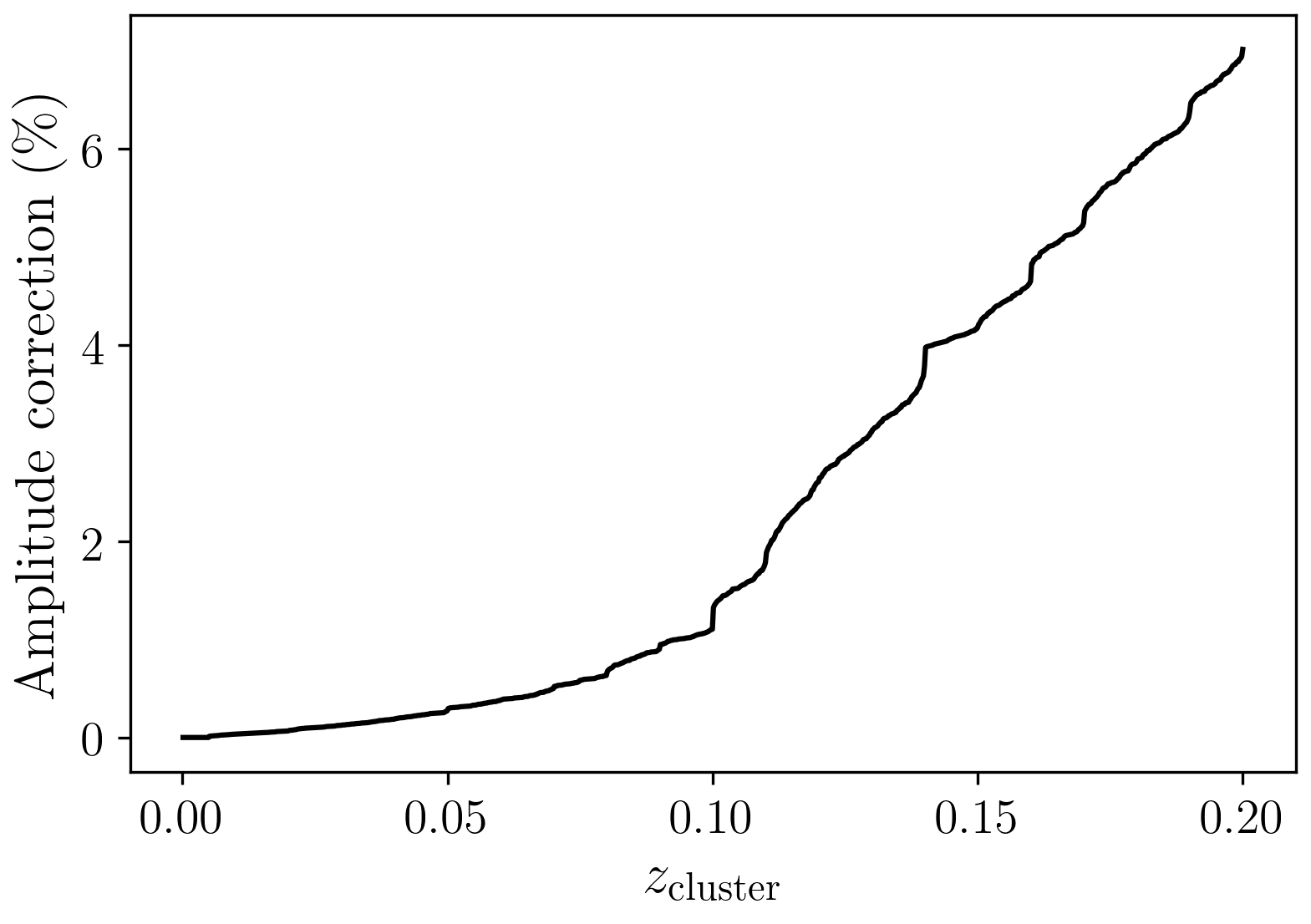}
    \caption{The demonstration of an amplitude correction by using the probability distribution function of DECaLS DR3 sources from the COSMOS 2015 catalog to correct the contaminating foreground galaxies.}
    \label{fig:amp_correct}
\end{figure}

By construction, foreground galaxies contain no lensing signal. Therefore on average, they decrease the amplitude of the lensing signal. To quantify the impact of the contamination by foreground galaxies, we compare the integrated lensing critical surface densities $\Sigma_{\text{cr}}$ defined in Equation~\ref{eq:int_pz}, 
when we include all galaxies $(z_{\text{source}} > 0)$,
and when we remove galaxies located in front of a cluster at redshift $z_{\text{cluster}}$.
In Figure \ref{fig:amp_correct}, we show the amplitude correction (A) defined as
\begin{eqnarray}
\label{eq:amp_correct}
\text{A (\%)} = 
\frac{\Sigma_{\text{cr}}(z_{\text{source}} > 0) - \Sigma_{\text{cr}} (z_{\text{source}} >  z_{\text{cluster}})}
{\Sigma_{\text{cr}} (z_{\text{source}} > 0) } \times 100 \% \, .
\end{eqnarray}
\noindent We find that the amplitude correction increases quickly with cluster redshift.
However, in the ideal case where we could remove foreground galaxies, the lensing
signal would only be increased by $\approx$ 4\%. This is marginal, given that our signal-to-noise at 
$R=5\ h^{-1}\text{Mpc}$ is about $\text{S/N}=10$. 
Nonetheless, since this amplitude correction increases quickly, it prevents us from studying
the mass-richness relation for clusters at redshift $z_{\text{cluster}}>0.2$.

\subsection{Weak lensing sample of CODEX clusters}\label{sec:codex}
\begin{figure}
	\includegraphics[width=\columnwidth]{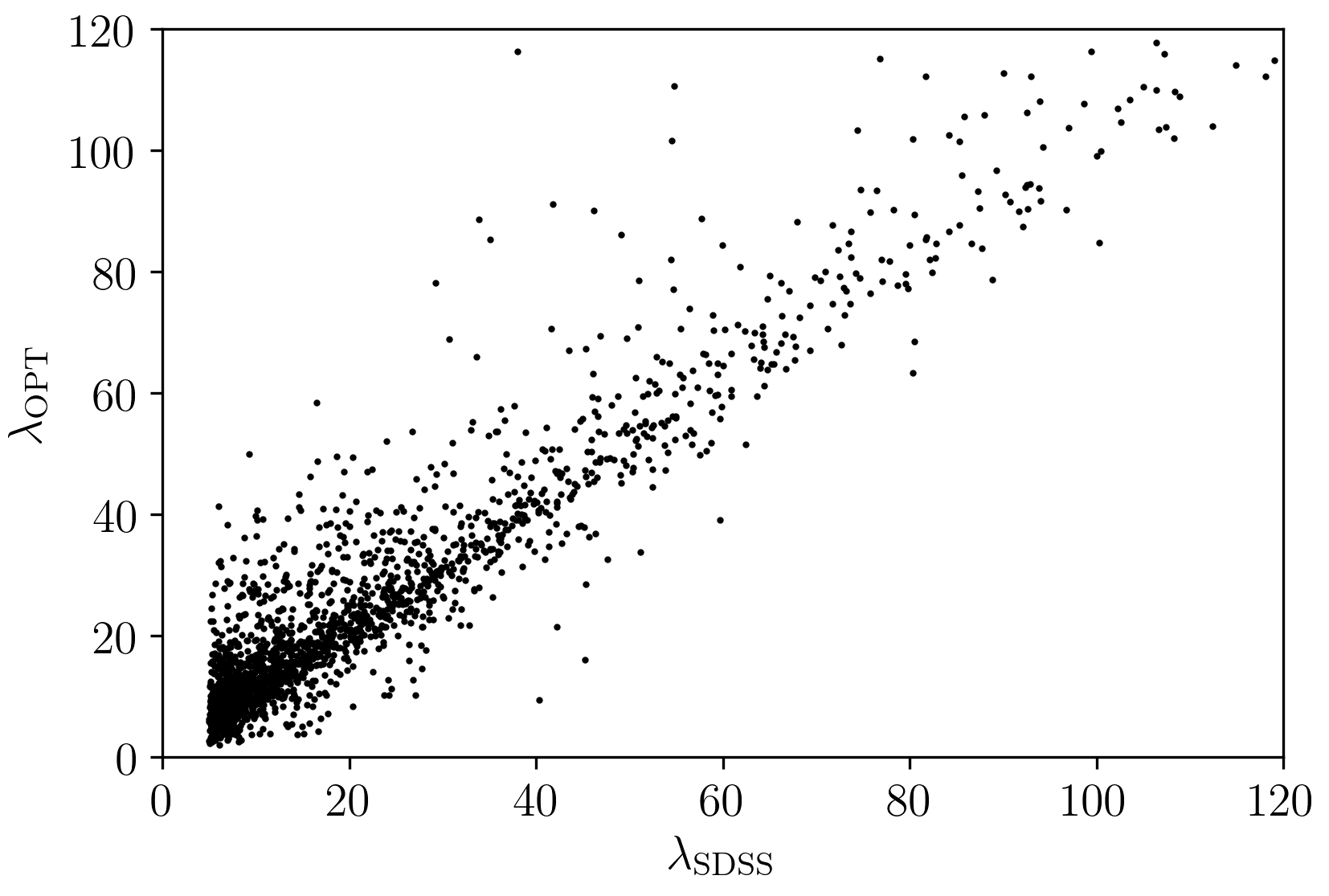}
    \caption{Comparison of the SDSS richnesses estimated by the redMaPPer algorithm within 3$^\prime$ from the X-ray center ($\lambda_{\text{SDSS}}$) and at the optically identified center ($\lambda_{\text{OPT}}$) of the CODEX clusters in the DECaLS DR3 survey footprint.}
    \label{fig:lsdd_lopt}
\end{figure}

The CODEX cluster catalog is constructed by identifying faint (selected down to a 4$\sigma$ significance of a source detection on spatial scales of 3-12 arcminutes) X-ray sources in the ROSAT All SKy Survey data \citep{voges1999} using redMaPPer algorithm \citep{rykoff2014}. The catalog contains a cluster redshift, a richness, a position of the optical center, 
selected according the redMaPPer rules, 
but constrained to be within 3$^\prime$ from the X-ray center, and a catalog of member galaxies. 
The set of redMaPPer parameters identified at the optical center are notated using the OPT suffix
($z_{\lambda, \text{OPT}}$, $\lambda_{\text{OPT}}$ and etc.). 
Cluster richness is evaluated at two positions, the X-ray ($\lambda_{\rm SDSS}$) and the optical ($\lambda_{\rm OPT}$). 
The results of the SDSS-IV spectroscopic follow-up using the catalog of member galaxies show a high degree of cluster confirmation at $z<0.4$ \citep{clerc2016}, which covers the redshift range used in this work (SDSS-IV; \citealp{dawson2016, blanton2017}). 
The results of a similar catalog construction in DES \citep{klein2018} show that for the redshift range considered here ($0.1 \leq z_{\rm cluster} \leq 0.2$), the use of a richness cut $\lambda \geq 20$ results in the catalog pure at 99\%. 
In Figure \ref{fig:lsdd_lopt}, we compare the two CODEX cluster richness estimates in the DECaLS DR3 footprint.
We observe a scatter of about 20\% at richness $\lambda_{\text{OPT}} = 20$, which decreases to $\sim 5$\% at $\lambda_{\text{OPT}} = 120$. In the following, we use redMaPPer parameters with the OPT suffix for the lensing analysis, because it yields higher signal-to-noise measurements.
 \citet{clerc2016} highlighted that $z_{\rm OPT}$ is systematically larger than the redshift
obtained in spectroscopy with the SPIDERS data, for  
clusters at redshift $z_\text{OPT} < 0.1$. This will affect the richness estimate, as the aperture for galaxy selection requires distance information taken from the redshift.
For this reason, we select clusters at redshift 0.1 $\leq$ $z$ $\leq$ 0.2, and optical richness 20 $\leq$ $\lambda_{\text{OPT}}$ $\leq$ 110. This way, we 
exclude low richness clusters, which might be contaminated by 
projected structures along the line-of-sight.

We cross-match the CODEX clusters with the area of the DECaLS DR3 data 
and select the galaxy clusters located in the DECaLS DR3 survey footprint.
The final subsample of CODEX clusters contains 279 clusters for our weak lensing analysis.
\begin{figure}
\centering
    \centering
	\includegraphics[width=\columnwidth]{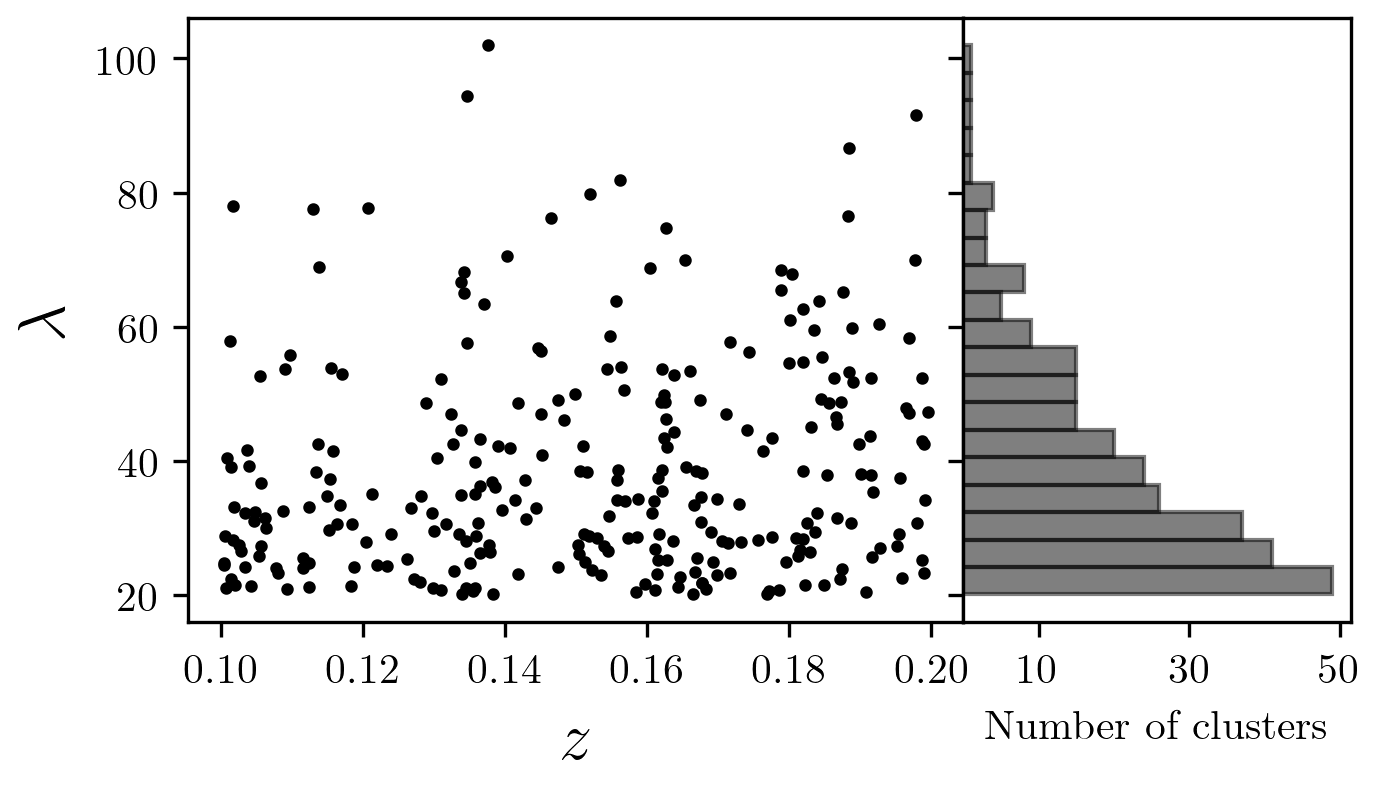}
    \caption{The distribution of the optical richness and cluster redshift of our weak lensing sample of CODEX clusters.}
    \label{fig:z_l}
\end{figure}
%
%

%%%%%%%%%%%%%%%%%%%%%%%%%%%%%%%%%%%%%%%%%%%%%%%%%
%%%%%%    Section III : Weak Lensing    %%%%%%%%%
%%%%%%%%%%%%%%%%%%%%%%%%%%%%%%%%%%%%%%%%%%%%%%%%%
\section{Modelisation}\label{sec:wl}
\subsection{Mass profile}
To ease comparison with previous works, we use the Navarro-Frenk-White profile (\citet{navarro1996}, hereafter NFW profile)
to model the dark matter halo of the galaxy cluster, complemented by a second term that accounts for the correlated matter distribution at large-scale.
In practice, we use the excess surface mass density as an estimator  \citep{mandelbaum2005}, which is given by two terms, 
\begin{eqnarray}
\label{eq:sigma_stacking}
\Delta \Sigma (R) = \Delta \Sigma_{\text{1h}}^{\text{NFW}}(R) + \Delta \Sigma_{\text{2h}}(R).
\end{eqnarray}
We measured the signal in the radius range $1.0 \leq R < 30.0 \, \text{Mpc}$, 
to avoid the mis-centering effect at small-scale and to maximize the constraints by increasing the number of bins at large-scale as discussed in Appendix \ref{sec:miscentering}.

\subsubsection{NFW profile}\label{sec:nfw_profile}
Numerical simulation is an effective tool to probe the large-scale structure problems.
\citet{navarro1996} used the N-body simulation in the cold dark matter cosmogony to investigate the density profile of the dark matter halos that obeys the following double power-law form,
\begin{eqnarray}
\label{eq:nfw_profile}
\rho_{\rm NFW}(r) = \frac{\delta_{c}\rho_{c}}{(r/r_{s})(1+r/r_{s})^2} \, ,
\end{eqnarray}
where $r$ is the distance from the cluster center in three dimensions, 
$r_{s}$ is the scale radius characteristic of the distribution of matter in the cluster, 
and $\delta_{c}$ is the linear overdensity threshold at which a halo collapses is defined as
\begin{eqnarray}
\label{eq:delta_c}
\delta_{c} = \frac{200}{3}\frac{c^3}{[\text{ln}(1 + c) - c/(1 + c)]} \, .
\end{eqnarray}
where $c$ is the dimensionless parameter which is called the concentration parameter.
%It is possible to define the dimensionless
%parameter which is called the concentration parameter, $c %\equiv r_{200} / r_{s}$; 
%$r_{200}$ is the radius inside which the mass density of %the halo is equal to 200 times the critical density.
The critical density of the universe can be written as
\begin{eqnarray}
\rho_{c} = \frac{3}{800 \pi} \frac{M_{\rm 200c}}{r_{200}^3},
\end{eqnarray}
where $M_{\rm 200c}$ is the cluster mass equal to the total mass inside that radius ($r_{200}$) or 200 times the critical density of the universe.
We assume that, on average, galaxy clusters are spherically symmetric, 
therefore the tangential shear of background sources at the redshift $z_{s}$ 
induced by the cluster lens at the redshift $z_{l}$ is given by
\begin{eqnarray}\label{eq:sigma_2}
\gamma_{t}(R) \equiv 
\frac{\overline{\Sigma}(<R) - \overline{\Sigma}(R)}{\Sigma_{\text{cr}}(z_{l}, z_{s})} 
= \frac{\Delta \Sigma (R)}{\Sigma_{\text{cr}}(z_{l}, z_{s})} \, ,
\end{eqnarray}
where $\overline{\Sigma}(R)$ is the mean surface density at the radius $R$, 
\begin{eqnarray}\label{eq:sigma_mean}
\overline{\Sigma}(R) = \overline{\rho}_{\rm m} \int_{-\infty}^{\infty} 
\xi_{\text{1h}}^{\rm NFW} \left( \sqrt{R^2 + \chi^2} \right) \text{d} \chi \, ,
\end{eqnarray}
and $\overline{\Sigma}(<R)$ is the mean surface density inside that radius,
\begin{eqnarray}\label{eq:sigma_mean2}
\overline{\Sigma}(<R) = \frac{2}{R^2} \int_{0}^{R}  \overline{\Sigma}(R^\prime) R^\prime \text{d} R^\prime \, ,
\end{eqnarray}
and $\Delta \Sigma (R)$ is called the excess surface mass density.
In Equation \ref{eq:sigma_mean}, $\xi_{\text{1h}}^{\rm NFW}$ is the halo-matter correlation function of the 1-halo term of the NFW profile,
\begin{eqnarray}
\xi_{\text{1h}}^{\rm NFW} = \frac{\rho_{\rm NFW}}{\overline{\rho}_{\rm m}} - 1 \, ,
\end{eqnarray}
where $\overline{\rho}_{\rm m} = \rho_{c} \Omega_{\rm m}$ is the mean matter density, $\rho_{c}$ is the critical density and $\Omega_{\rm m}$ is the matter density of the universe (e.g. \citealp{melchior2017, mcclintock2019apj}).

The relation between the tangential shear and the excess mass density can be derived 
from the critical surface mass density of the lens $\Sigma_{\text{cr}}(z_{l}, z_{s})$ 
in comoving coordinates as 
\begin{eqnarray}\label{eq:sigma_crit}
\Sigma_{\text{cr}}(z_{l}, z_{s}) = \frac{c^2}{4 \pi G }
\frac{D_{A}(z_{s})}{ D_{A}(z_{l}) D_{A}(z_{l}, z_{s})}
\frac{1}{(1 + z_{l})^{2}},
\end{eqnarray}
where $D_{A}(z_{l})$, $D_{A}(z_{s})$ and $D_{A}(z_{l}, z_{s})$ are the angular diameter distances to the lens, sources and between the lens and sources respectively.
In practice, instead of computing the critical surface mass density,
we estimate an effective critical surface density
(\citealp{okabe2014, mcclintock2019}) by integrating over the source distribution,
\begin{eqnarray}\label{eq:int_pz}
\left < \Sigma_{\text{cr}}^{-1}(z_{l}) \right > = \frac
{\int  P(z_{\text{s}}) \, \Sigma_{\text{cr}}^{-1}(z_{l}, z_{s}) \, \text{d} z_{s}}
{\int P(z_{\text{s}}) \, \text{d} z_{s}} \, ,
\end{eqnarray}
where $P(z_{\text{s}})$ is the probability distribution function of source galaxies
that is defined in Equation \ref{eq:prob_zs}.
Thus, we can finally relate the excess surface mass density to the weak lensing measurement 
 by 
\begin{eqnarray}\label{eq:final_nfw}
\Delta \Sigma_{\text{1h}}^{\text{NFW}}(R) = 
\gamma_{t}(R) \left < \Sigma_{\text{cr}}^{-1}(z_{l}) \right >^{-1} \, .
\end{eqnarray}
%where $(1 + m)$ is the correction factor defined in Equation \ref{eq:oneplusm}.

\subsubsection{The 2-halo term}
Galaxy clusters reside at the node of the cosmic web, with dense filament and massive nearby halos.
The mass measurement of a galaxy cluster 
is not perfectly consistent 
with the underlying matter density field, 
suggesting that the galaxies are not exactly tracking the underlying mass distribution
(see e.g. \citealp{dekel1999, jullo2012}).
The matter distribution around galaxy clusters shapes
the mass profile at large scale (\citealp{hoekstra2011, jauzac2012}).
We can express the excess surface mass density of the second dark matter halo by
\begin{eqnarray}\label{eq:excess_2ndhalo}
\Delta \Sigma_{\text{2h}} (R) = \overline{\Sigma}_{\text{2h}} (<R) - \overline{\Sigma}_{\text{2h}} (R) \, .
\end{eqnarray}
The excess mass density at the projected radius $(R)$ is defined by
\begin{eqnarray}
\overline{\Sigma}_{\text{2h}} (R) = 2\rho_{c,0} \Omega_{\rm m, 0} \int_{0}^{\infty} 
\xi_{\text{2h}} \left( \sqrt{R^2 + \chi^2} \right) \text{d} \chi \, ,
\end{eqnarray}
where $\xi_{\text{2h}}(r)$ is the galaxy-matter cross-correlation function, $\rho_{c,0}$ is the critical density, and $\Omega_{\rm m, 0}$ is the matter density of the universe at the present time.
This function is obtained by multiplying the non-linear matter correlation function $\xi_{\rm nl}(r)$ with the halo bias $b(M, z)$ given by
\begin{eqnarray}
\xi_{\text{2h}}(r) = b(M, z) \, \xi_{\rm nl}(r)\,  \zeta(r) \, ,
\end{eqnarray}
where $\zeta(r)$ is the scale dependence of halo bias which is more significant 
at the radius $R \leqslant 3$ $h^{-1}$ Mpc, described in \citet{tinker2005}.
In this study, we use the halo bias prescription derived from cosmological simulations
by \citet{tinker2010}.
The nonlinear matter correlation function is the Fourier transform of the nonlinear matter power spectrum $P_{\text{\rm nl}}$,
\begin{eqnarray}
\xi_{\rm nl}(r) = \frac{1}{2\pi} \, \int k^{2} \, P_{\rm nl}(k) \, j_0(kr) \, \text{d}k \, ,
\end{eqnarray}
where $j_0(kr)$ is the zeroth order of spherical Bessel function of the first kind.
We use the revised halo fit model \citep{takahashi2012} 
and the Code for Anisotropies in the Microwave Background (CAMB) program \citep{camb2011}
to compute the nonlinear matter power spectrum
by evaluating the mean redshift of stacked clusters.

\subsection{Concentration-Mass relation}
The NFW profile is described by two parameters; the concentration ($c$) and the total mass ($M_{\rm 200c}$).
The results from the numerical simulations show that the concentration and the total mass ($c$-$M$) are related.
Data from observations are also used to test the $c$-$M$ relation \citep[e.g.][]{okabe2010}.
The relation between the concentration and the total mass is an important tool
to test the cosmological models and the physical processes in a galaxy cluster.
There are many studies in different datasets to characterize this relation.
\citet{duffy2008}, for example, studied the $c$-$M$ relation of the NFW profile using N-body simulations
in the Wilkinson Microwave Anisotropy Probe year 5 (WMAP5) cosmology, and found an additional dependency between halo masses and redshift, best modeled by a power-law.
The results show that the concentration decreases as a function of the total mass and redshift.
Later, \citet{dutton2014} studied the $c$-$M$ relation in the Planck cosmology.
They found that the $c$-$M$ relation is higher than the study in the WMAP5 cosmology.
They provided the following expression for the concentration mass relation, defined as a power-law
\begin{eqnarray}\label{eq:cm_dutton1}
\text{log}_{10} c_{\rm 200c} = a + b \, \text{log}_{10} (M_{\rm 200c} / [10^{12} h^{-1} M_\odot]),
\end{eqnarray}
where $a$ and $b$ are given by
\begin{eqnarray}\label{eq:cm_dutton2}
a = 0.520 + (0.905 - 0.520) \, \text{exp}(-0.617z^{1.21}),
\end{eqnarray}
\begin{eqnarray}\label{eq:cm_dutton3}
b = -0.101 + 0.026z.
\end{eqnarray}
For this analysis, the signal-to-noise of our measurement is quite low.
Consequently, we use this relation to reduce the number of free parameters in our model.
%There are small biases from the different relations. 
However, we note that this $c$-$M$ relation determined from a complete set of simulated clusters, 
might differ from one of our samples, especially at low mass, where we might 
miss some clusters. 
Nonetheless, the previous study from \citet{cibirka2017} on the CODEX clusters shows an excellent agreement 
between the simulation from \citet{dutton2014} and observations.

\section{Analysis}\label{sec:analysis}
\subsection{Weak lensing estimator}\label{sec:wl_estimator}
As we discussed, the weak lensing analysis requires deep imaging data.
The density of background galaxies of the DECaLS DR3 is not sufficient to measure the mass of individual clusters.
Therefore, we turn to the stacking technique to recover the weak lensing signal. 
The stacking technique is the method in which the galaxy clusters 
and their measured radial density profile
can be combined according to observed properties (e.g. richness or X-ray luminosity).
It helps to enhance the S/N of the lensing profile when we cannot recover the lensing signal of individual clusters.
The stacked surface mass density profile can be written in terms of the summation of the tangential shear over $N$ background galaxies $i$ 
that are found within an annular region at radius $R$ around the lenses as
\begin{eqnarray}
	 \Delta \Sigma (R) = C(R) \frac{\sum_{\text{i} = 1}^N w_{\text{i}} \gamma_{ \text{t,i}} (R)
	 \left\langle \Sigma_{\text{cr}}^{-1} \right\rangle^{-1}}
	 {\sum_{\text{i} = 1}^N (1 + m) w_{\text{i}}}, 
\end{eqnarray}
where $C(R)$ is called a boost factor and $w_{\text{i}}$ is the weight chosen to minimize the variance of a shear estimator.
We include a boost factor to account for contamination in our lensing source sample by galaxies that might be associated with the clusters (\citealp[see e.g.][]{fischer2000, simet2017}). 
This effect is scale dependent, in contrast with the critical density rescaling 
$ \left\langle \Sigma_{\text{cr}}^{-1} \right\rangle^{-1}$, due to the lack of photometric redshift information.
For lensing sources with index $i$, around $N$ lenses  with index $j$, and sources with index $k$,
around $N_{\text{rand}}$ random points with index $l$, the correction factor is given by
\begin{eqnarray}
	C(R) = \frac{N_{\text{rand}}}{N} \frac{\sum_\text{i,j} w_\text{i, j}}{\sum_\text{k, l} w_\text{k, l}}.
\end{eqnarray}
Compared to previous works, we adapt the method to estimate the correction factor 
by drawing 10 random positions in a 1.5 degrees aperture from the cluster centers, instead of drawing random points in the full survey footprint. This is justified by the fact that cluster density is very low, and our survey footprint is very irregular. We found that the variation of the boost factor profile is less than 5\%  within the size of the aperture. More detail is given in Appendix \ref{sec:dilution_galaxies}. Overall, the boost factor effect is less than 1\% at $R > 1 $ Mpc, where we perform our analysis.
\subsection{Error estimation}\label{sec:error_estimation}
\subsubsection{Stacked samples covariance}
\label{sec:stacked_samples_covariance}
There are many effects which cause statistical errors in the measurements, such as
the intrinsic shape of source galaxies, the number of lens-source pairs 
and the fluctuations of the large-scale structure along the line-of-sight \citep[e.g.][]{shirasaki2017}.
We use the Jackknife technique detailed in the following procedure to estimate the statistical errors for each of the three stacked cluster groups:
(i) Randomly draw 10 positions within a 1.5 degrees aperture from true cluster positions
(ii) Employ the delete-1 Jackknife technique in each realization by removing one cluster
from the stacked profile and average the lensing profiles for true and random clusters 
(iii) Repeat the measurement for each Jackknife configuration.
The covariance matrix of stacked clusters for the Jackknife technique can be written as
\begin{eqnarray} \label{eq:cov_jk}
	\mathbf{C} \equiv C_{ij} = \frac{N_{\text{JK}} - 1}{N_{\text{JK}}} \sum_{m = 1}^{N_{\text{JK}}}
	\left( \Delta\Sigma_{i}^{m} - \overline{\Delta\Sigma}_{i} \right)
	\left( \Delta\Sigma_{j}^{m} - \overline{\Delta\Sigma}_{j} \right) \, .
\end{eqnarray}
where $i$ and $j$ indicate the radial bins, $\Delta\Sigma^{m}$ is the excess surface mass density for each of the $m$-th Jackknife configuration.
In equation above, the mean excess surface mass density of 
the Jackknife configurations is defined by
\begin{eqnarray}
\overline{\Delta\Sigma} \equiv \frac{1}{N_{\text{JK}}}
\sum_{m = 1}^{N_{\text{JK}}} \left ( 
\overline{\Delta\Sigma}_{\text{true}}^{m}   -  \overline{\Delta\Sigma}_{\text{random}}^{m} \right ) \, ,
\end{eqnarray}
where $\overline{\Delta\Sigma}_{\text{true}}$ is the mean excess surface mass density of true cluster positions and 
$\overline{\Delta\Sigma}_{\text{random}}$
is the mean excess surface mass density of random positions
in each Jackknife configuration.
Notwithstanding, the covariance matrix of the Jackknife is underestimated due to the noise level, we therefore multiply the inverse covariance matrix ($\mathbf{C}^{-1}$) by the Hartlap factor; $H = (N_{\rm JK} - p - 2)/(N_{\rm JK} - 1)$, where $N_{\rm JK}$ is the number of Jackknife configurations and $p$ is the number of measured radial bins \citep{hartlap2007}.

\subsubsection{Individual cluster covariance}
\label{sec:individual_cluster_covariance}
In addition, we measure the covariance of the excess surface mass density profile for a typical cluster in our sample, to be used later in Section \ref{sec:mass_richness_relation}. 
To this end, we estimate the covariance matrix by 
\begin{eqnarray}
\mathbf{C} = \frac{1}{N - 1} \sum_{i = 1}^{N} 
(\mathbf{\Delta \Sigma}_{i} - \overline{\mathbf{\Delta \Sigma}}) \cdot 
(\mathbf{\Delta \Sigma}_{i} - \overline{\mathbf{\Delta \Sigma}})^{\rm T} \, ,
\label{eq:cov2}
\end{eqnarray}
where $\mathbf{\Delta \Sigma}$ is a two-dimensional matrix, in which
rows and columns correspond to the radial bins, the individual cluster measurements respectively.  The vector $\overline{\mathbf{\Delta \Sigma}}$ is the average of the excess surface mass density profile of all clusters \citep[e.g.][]{mcclintock2019}.
We assume that cluster measurements are uncorrelated, and
therefore, that the covariance matrix is diagonal. This statement is supported by the fact that off-diagonal terms derived from Equation \ref{eq:cov2} are extremely noisy.

\subsection{MCMC method}
The Markov Chain Monte Carlo (MCMC) method is a resampling technique 
based on probability distributions.
We use the MCMC method from the emcee package written in Python \citep{daniel2013}
to fit our model to the measured density profiles.
We set the number of walkers, burn-in(s) and production samples
to 48, 1000 and 10000 respectively.
In our analysis, we assume that the log-likelihood function 
of the observed data ($\Delta\Sigma^{\text{obs}}$) is given by
\begin{eqnarray}
\text{ln} \mathcal{L} (\Delta\Sigma^{\text{obs}} | \Delta\Sigma^{\text{model}}) \propto -\frac{1}{2} \mathbf{D}^{\rm T} \mathbf{C}^{-1} \mathbf{D} \,
\end{eqnarray}
where $\mathbf{D}$ = 
($\mathbf{\Delta\Sigma}^{\text{obs}} - \mathbf{\Delta\Sigma}^{\text{model}}$)
is the differences between the observed and modeled excess surface mass density, and $\mathbf{C}$ is the covariance matrix described in Section \ref{sec:error_estimation} \citep[e.g.][]{mcclintock2019}.
\begin{figure*}
	\includegraphics[width=\textwidth]{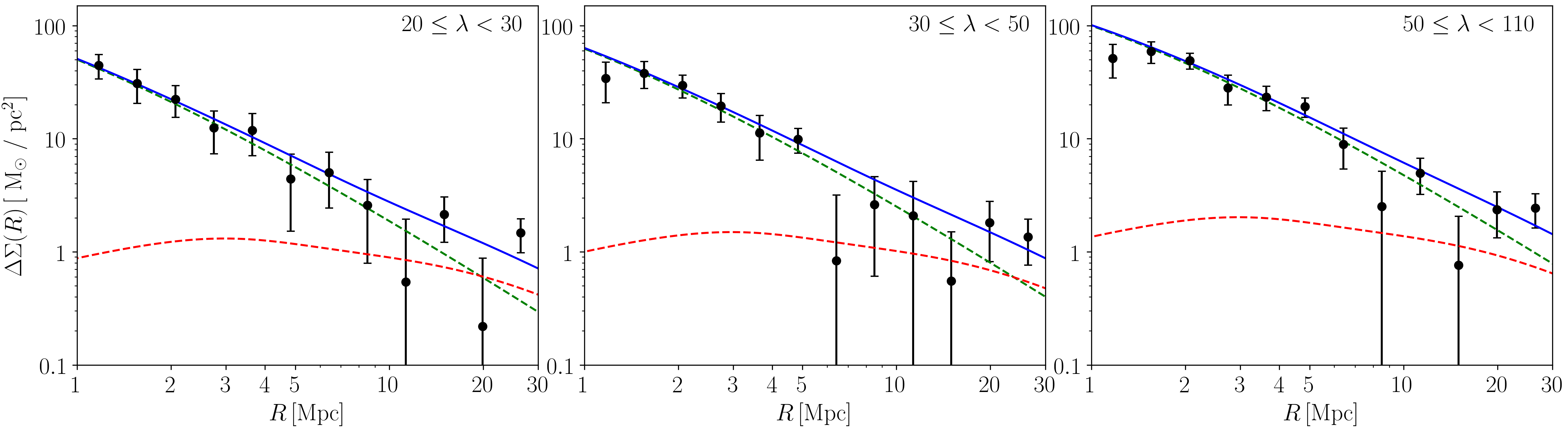}
    \caption{The excess surface mass density profiles of CODEX clusters in three richness groups: 
20 - 30 , 30 - 50 and 50 - 110. The blue solid line shows the theoretical profile defined in Equation \ref{eq:sigma_stacking} with the first halo term in green and second halo term in red dashed lines. The error bars show the square root of the diagonal terms of Jackknife covariance matrix.}
    \label{fig:excessprofile}
\end{figure*}
\begin{table*}
\caption{The results from the mass measurements in three richness bins. 
The excess surface mass density profiles have been fitted with the theoretical model in two regions; 
the inner region ($1.0 \leq R < 5.5$ Mpc) modeled with the NFW profile only
($\Delta \Sigma^{\text{NFW}}_{\text{1h}}$),
and the whole region ($1.0 \leq R < 30.0$ Mpc) modeled with the two halo terms 
($\Delta \Sigma^{\text{NFW}}_{\text{1h}}$ + $\Delta \Sigma_{\text{2h}}$). The chi-square goodness of fit test and the degree of freedom (dof) reveal the best fits between observed data and the theoretical profile.} 
\label{tab:table_cluster}
\begin{tabular}{cccccccc}
\hline
Richness Group & N$_{\text{cluster}}$ & $z_{\text{mean}}$ & $\lambda_{\text{opt, mean}}$ 
& $M_{\rm 200c}^{\text{1h}}$ ($10^{14}$ M$_{\odot}$)
& $\chi^2_{\text{1h}}$ / dof
& $M_{\rm 200c}^{\text{1h + 2h}}$ ($10^{14}$ M$_{\odot}$)
& $\chi^2_{\text{1h + 2h}}$ / dof\\
\hline
\rule{0pt}{10pt}
[20, 30) &  109  & 0.148 & 24.83 $\pm$ 2.98 & $2.19^{+0.39}_{-0.37}$ & 1.17 / 4 & $1.91^{+0.27}_{-0.27}$ & 6.91 / 10\\
\rule{0pt}{12pt}
[30, 50) & 110 & 0.152 & 38.90 $\pm$ 5.86 & $2.98^{+0.44}_{-0.42}$ & 3.70 / 4 &  $2.37^{+0.32}_{-0.32}$ & 9.58 / 10\\
\rule{0pt}{12pt}
[50, 110) & 60 & 0.158 & 63.86 $\pm$ 11.65 & $5.91^{+0.82}_{-0.79}$ & 7.15 / 4 &  $5.02^{+0.52}_{-0.50}$ & 13.52 / 10\\
\hline
\end{tabular}
\end{table*}

\section{Results}\label{sec:results}
\subsection{Samples of CODEX clusters}
\label{sec:results_samplecodex}
We split our CODEX clusters sample into three richness groups; 
$\lambda =$ 20 - 30, 30 - 50 and 50 - 110.
In each group, we stack the profiles and fit them
with the model defined in Equation \ref{eq:sigma_stacking} 
in the radial range $1.0 \leq R < 30.0$ Mpc.
We use the MCMC method to estimate confidence intervals and set the uniform prior on the cluster mass parameter, $M_{\rm 200c}$ = [$10^{13}$, $10^{15}$].

We find that the excess surface mass density at large radius (eg., $R$ > 5 Mpc) has a large scatter, which might affect our mass estimates.
Therefore, we repeat the mass measurement in the inner region only ($1.0 \leq R < 5.5$ Mpc), and fit only the first halo term ($\Delta \Sigma_{\rm 1h}^{\text{NFW}}$).
We report the best-fit values associated 
with their $1\sigma$ uncertainties in Table \ref{tab:table_cluster}.
The results show that the mass of CODEX clusters 
increases as a function of the optical richness.
Furthermore, we find consistent results between two fitting procedures, suggesting that the \citet{tinker2010} bias model is consistent with the 1h-term mass estimate.
In Figure \ref{fig:excessprofile}, we plot the excess surface mass density profile from the weak lensing measurement (black dots)
and the solid line shows our theoretical model.

\subsection{Mass-Richness Relation of the CODEX clusters} \label{sec:mass_richness_relation}
In this section, we assess the robustness of our estimated mass-richness relation 
with an alternative but equivalent method. Instead of adjusting a theoretical 
lensing signal to cluster density profiles distributed in bins of richness, we 
adjust a lensing signal to each profile based on the cluster richness and a parametrized mass-richness relation. 
We define the relation between the expected mass, richness and cluster redshift as
\begin{eqnarray}\label{eq:mass-richness}
\left\langle M_{\rm 200c} | \lambda \right\rangle = 
M_{0} \left( \frac{\lambda}{\lambda_{0}} \right) ^ {F_{\lambda}},
\end{eqnarray}
where $M_{0}$ and $F_{\lambda}$ are the free parameters of this relation.
We set the pivot values $\lambda_{0} = 40$, 
close to the mean optical richness of CODEX cluster samples.
Our data is not sensitive to more complicated Mass-richness models, we therefore exclude the redshift evolution term from this relation,  
because the narrow range in redshift of our clusters sample does not provide enough constraint to estimate the redshift scaling index with precision.
In addition, previous studies on this scaling relation 
found a minimal redshift evolution \citep[e.g.][]{mcclintock2019}.

\begin{figure}\centering
	\includegraphics[width=0.9\columnwidth]{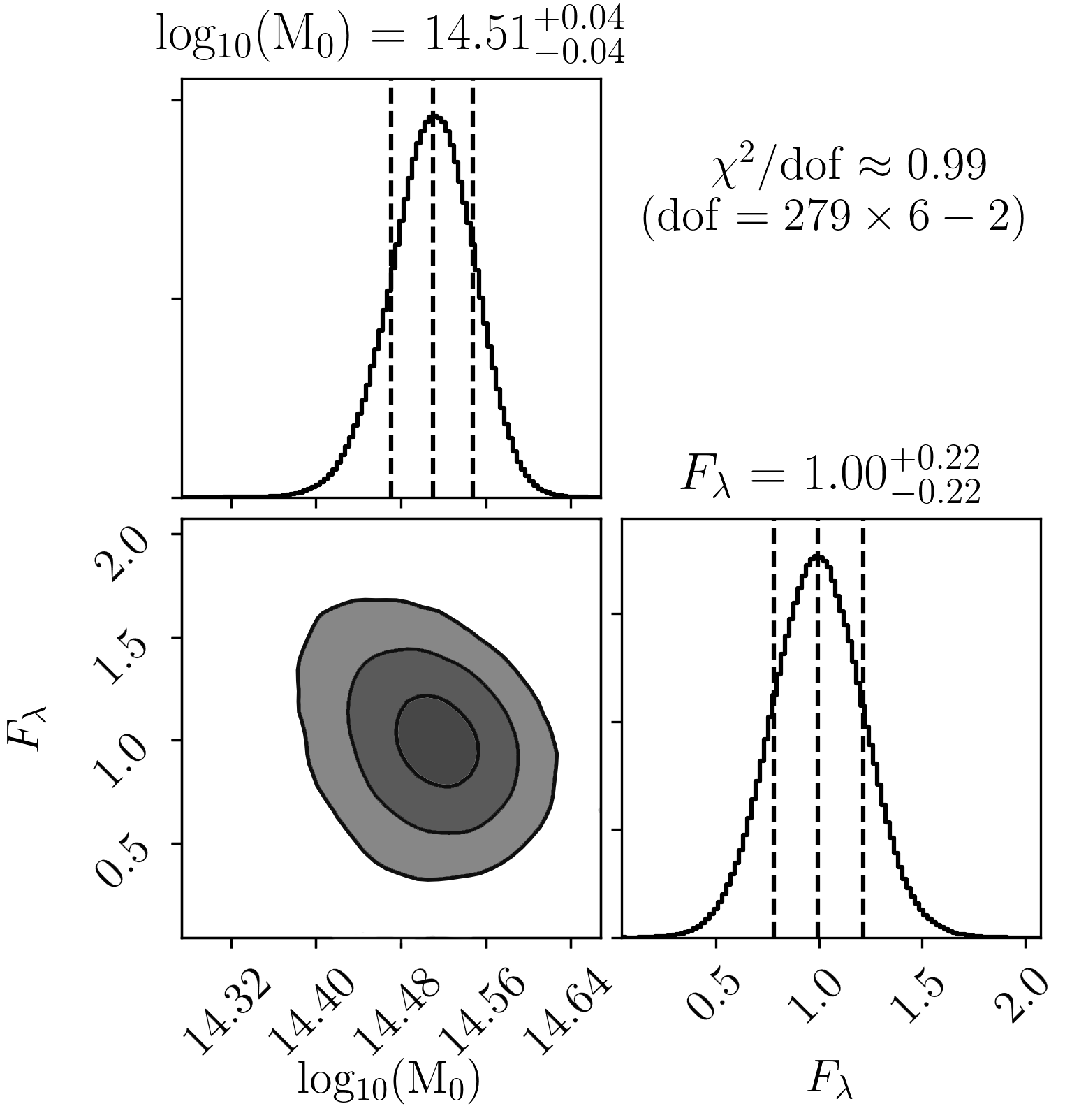}
    \caption{Posterior distributions of the free parameters in our Mass-Richness relation for our CODEX clusters sample, estimated with the MCMC analysis described in Section~\ref{sec:mass_richness_relation}. Vertical dashed lines indicate the 68\% confidence levels and contour lines correspond to 1$\sigma$, 2$\sigma$, and 3$\sigma$ confidence levels. }
    \label{fig:mcmc}
\end{figure}

We use the MCMC method with the following priors on the mass-richness relation parameters, log$_{10} M_{0}$ = [13.0, 15.0] and $F_{\lambda}$ = [-10.0, 10.0] with the initial values, log$_{10} M_{0}$ = 14.0 and $F_{\lambda}$ = 0.1. 
For consistency with previous measurements, we restrict our analysis to the inner region ($1.0 \leq R < 5.5$ Mpc) 
and fit each cluster with the NFW profile as described in Equation \ref{eq:final_nfw}.
In Figure \ref{fig:mcmc}, we plot the best fit estimate of the mass-richness relation parameters log$_{10} M_{0}$ and $F_\lambda$. 
We estimate the mean mass of our CODEX cluster sample,
\begin{eqnarray}
M_{0} = 3.24_{-0.27}^{+0.29} \times 10^{14} \, \text{M}_\odot \, ,
\end{eqnarray}
and the richness scaling index 
\begin{eqnarray}
F_{\lambda} = 1.00 ^{+0.22}_{-0.22} \, .
\end{eqnarray}
Clearly, the cluster mass and richness of our samples are strongly correlated, and consistent with a power-law relation 
given that the reduced chi-squared is $\chi^{2}/\rm{dof} \approx 0.99$.
We also fit the two halo terms model in the radial range $1.0 \leq R < 30.0$ Mpc and obtain the mean mass of the CODEX clusters,
$M_{0} = 2.79_{-0.22}^{+0.24} \times 10^{14} \, \text{M}_\odot$, and the richness scaling index, $F_{\lambda} = 0.98 ^{+0.20}_{-0.20}$.

Finally, in Figure \ref{fig:scaling2}, we compare our estimated masses in bins of richness with the best fit mass-richness relation found above 
and find a good agreement between both measurements. 
This gives us confidence that our two methods are
equivalent and our mass-richness relation in Equation \ref{eq:mass-richness} is a good model of the data.
\begin{figure}
\includegraphics[width=\columnwidth]{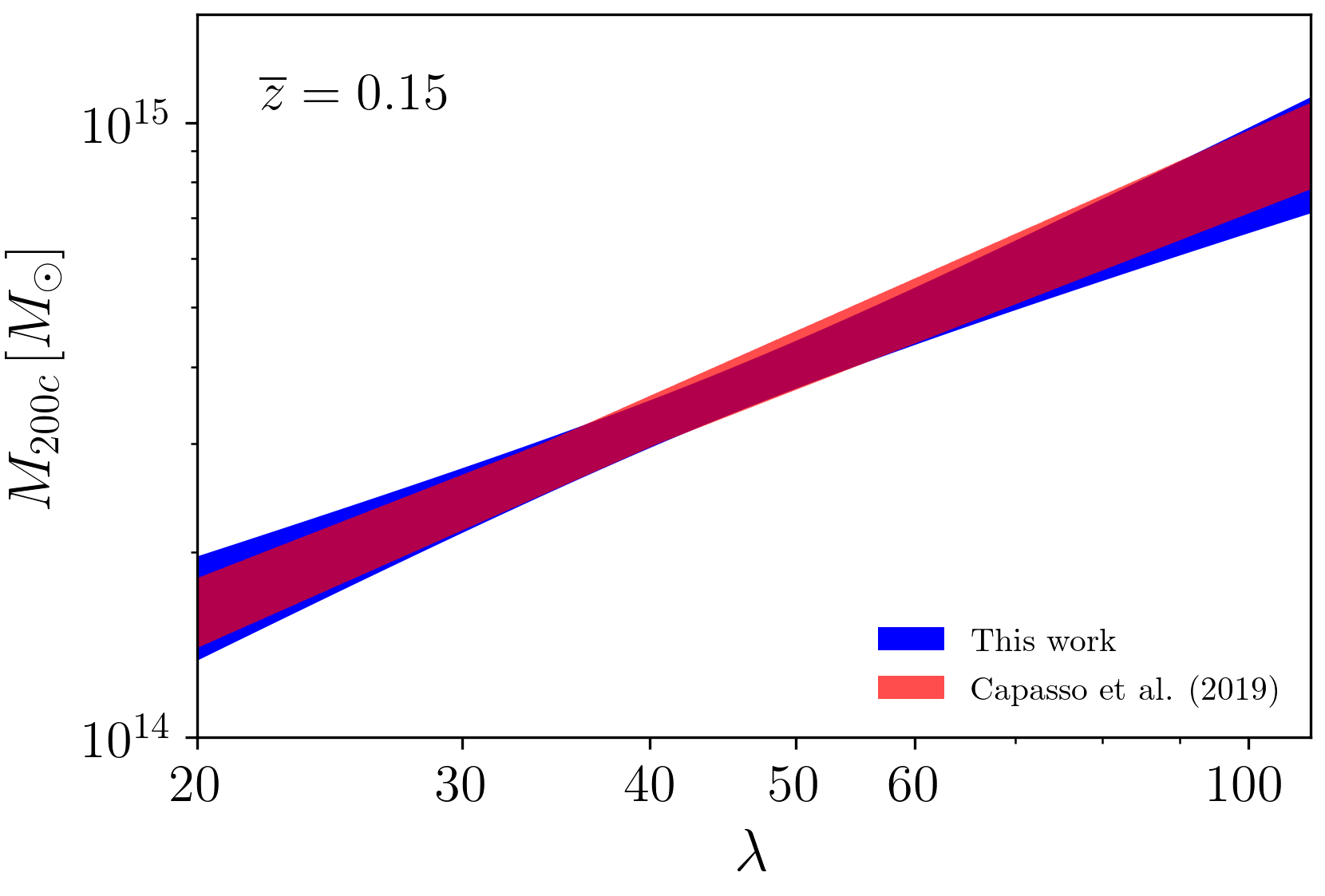}
\caption{Comparison of scaling relations determined with weak lensing and dynamical analysis for the CODEX cluster sample.}
\label{fig:scaling1}
\end{figure}

\subsubsection{Accuracy of richness estimates}
Our determination of the mass-richness relation relies on an accurate estimate of 
the cluster richness. However, a biased estimate of the richness may lead to
a biased relation \citep{rozo2009}. We must investigate how the observable richness is actually tracing the cluster mass. 

In our analysis, we use the optical richness as a mass tracing parameter.
In \citet{capasso2019}, the authors determine a correction factor ($\eta$) on the 
optical richness parameter of the CODEX clusters. This factor 
corrects the richness of clusters that are at the limit of the observational capabilities. This factor also enters in the calculation of the Poisson variance, and must be taken into account in the fit of the mass-richness relation. 
Another term is the intrinsic scatter $\sigma_{\text{int}}$ in the richness \citep[e.g.][]{simet2017}. It is constant for all clusters, whatever their richness is. Therefore, 
we can express the variance of the mass-richness relation as
\begin{eqnarray}\label{eq:scatter_eq1}
\text{Var}(\text{ln M} | \lambda) = \frac{\alpha^{2}}{\lambda}
+ \sigma_{\text{int}}^2 \, .
\end{eqnarray}
The introduction of this variance in the fit will down-weight both low and high 
richness clusters, either because the precision on low richness values is Poisson 
noise dominated or because the precision on high richness values is intrinsically 
limited.

We include $\alpha$ and $\sigma_{\text{int}}$ as free parameters in our model and use the MCMC technique to find their best fit values.
We find that the low signal-to-noise in the shear signal of each cluster prevents us from determining these values with good precision.
In order to reach a 7\% precision as in \citet{simet2017},
we would need roughly four times more clusters.
Therefore, in order to test their effect, 
we set $\alpha$ = 1 and $\sigma_{\text{int}} = 0.22$. Doing so, we assume that our clusters at low redshift are not significantly affected by observational considerations. 
Also, these values were already determined in \citet{capasso2019} for the CODEX clusters as well.
Taking into account these two variance terms, we find a mean cluster mass 
\begin{eqnarray}\label{eq:scatter_eq2}
    M_{0} = 3.12_{-0.26}^{+0.28} 
    \times 10^{14} \, \text{M}_\odot \, ,
\end{eqnarray}
and a richness scaling index
\begin{eqnarray}\label{eq:scatter_eq3}
    F_{\lambda}= 1.01_{-0.21}^{+0.22} \, .
\end{eqnarray}
These results are in statistical agreement with the results determined without 
taking into account the variance terms, which means that the richness estimates in our low redshift cluster sample are robust.
\begin{figure}
\includegraphics[width=\columnwidth]{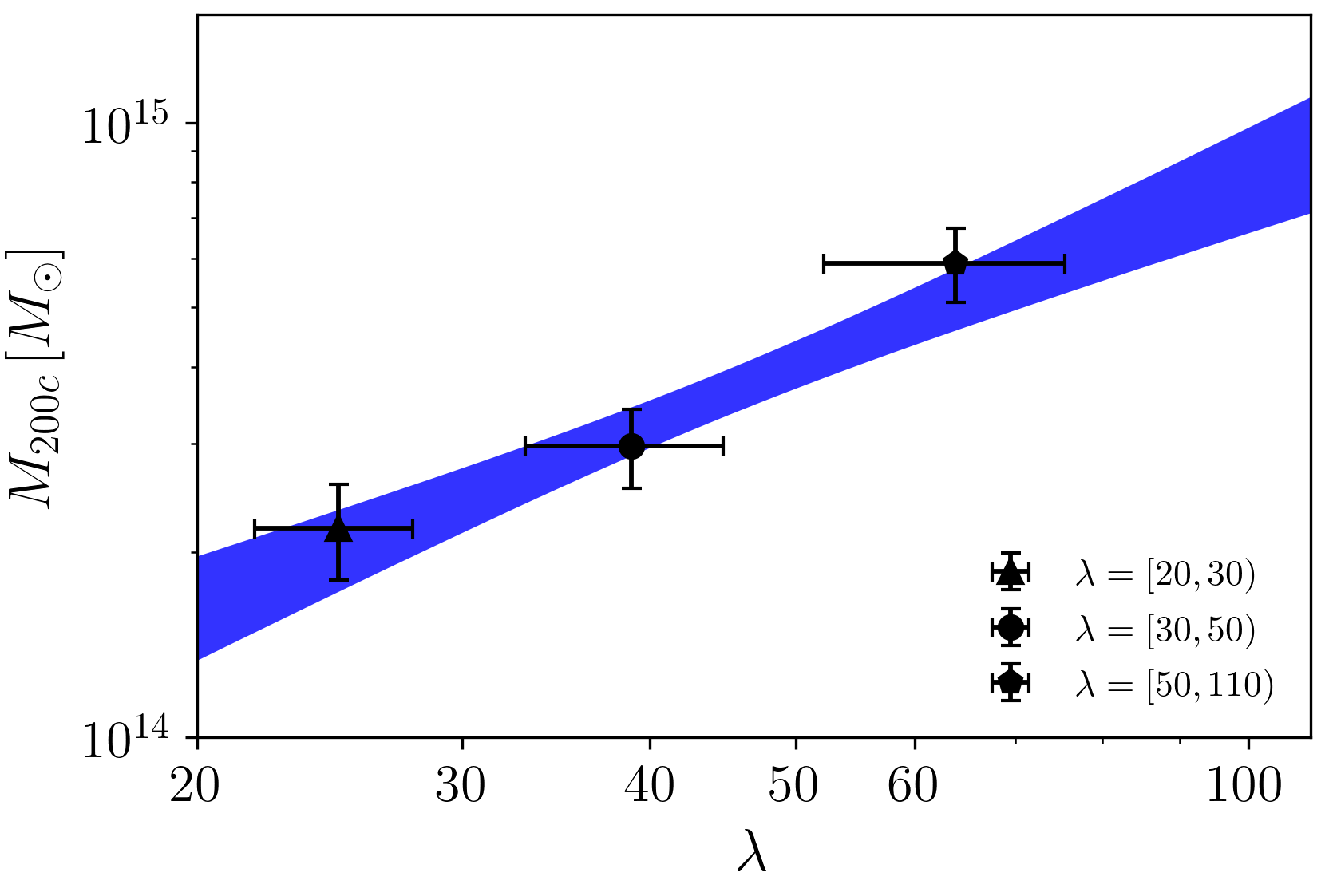}
\caption{
The scaling relation of CODEX clusters (blue shaded line) computed with individual cluster same as in Figure \ref{fig:scaling1} compared to the stacked galaxy clusters in three richness groups. The results from both methods are consistent with each other.}
\label{fig:scaling2}
\end{figure}

\subsubsection{CODEX clusters comparison with redMaPPer clusters}
Differences in the selection functions of the redMaPPer and CODEX clusters have already been detailed in Section~\ref{sec:codex}.
However, it is interesting to compare the mass-richness relation determined with our CODEX sample and with the redMaPPer clusters. 
As a reference for the redMaPPer clusters sample, we use the results from DES Year 1 analysis in \citet{mcclintock2019}.
They constrain the mass-richness relation for redMaPPer galaxy clusters in the redshift range $0.2 \leqslant z \leqslant 0.65$ and with richness values $\lambda \geqslant 20$. They obtain a richness scaling index $F_{\lambda} = 1.356 \, \pm \, 0.051 \, \text{(stat)}$ 
and a redshift scaling index $G_{z} = -0.30 \, \pm \, 0.30 \, \text{(stat)}$.
In comparison, our estimate of the richness scaling index is smaller. This suggests that at low richness, CODEX clusters are more massive than redMaPPer clusters. We 
attribute this difference to the completeness and selection function of the CODEX 
clusters, which have been confirmed with X-ray observations, and therefore are less
contaminated by possible line-of-sight confusions.

\subsection{Weak Lensing and Dynamical Analysis}\label{sec:wl-dynamic}
As detailed in the introduction, there are many methods to estimate cluster masses.
In this section, we compare the weak lensing masses from our analysis with
the dynamical masses determined from the velocity dispersion of galaxy cluster members.
Usually, dynamical analyses are based on the assumption that galaxy clusters 
are in a dynamical equilibrium state.
The cross study between weak lensing analysis and dynamical analysis can be used to calibrate a cluster mass and study the dynamical equilibrium state in galaxy clusters.
However, within the limit of the survey, 
we only compare the scaling relation between both methods 
in order to track the cluster mass as a function of richness.

We compare our measurements with the dynamical analysis performed by \citet{capasso2019}, assuming the clusters observed by the SPIDERS program in spectroscopy constitute a random subsample of the parent CODEX cluster sample used in our analysis. 
They measured a sample of 428 galaxy clusters up to redshift z $\sim$ 0.66 and model the scaling relation as
$\lambda \propto A_{\lambda} \, M_{\rm 200c}^{B_{\lambda}} \, (1 + z)^{\gamma_{\lambda}}$, where $A_{\lambda} = 38.6_{-4.1}^{+3.1}$, 
$B_{\lambda} = 0.99_{-0.07}^{+0.06}$ and $\gamma_{\lambda} = -1.13_{-0.34}^{+0.32}$.
In Figure \ref{fig:scaling1}, we plot the weak-lensing and the dynamically-based scaling relations of the CODEX clusters. For both relations, we set the mean cluster redshift $\overline{z} = 0.15$. The width of the shaded areas corresponds to the 1$\sigma$ confidence level.
We found an excellent agreement between the two scaling relations. 
This result supports our claim that the SPIDER clusters constitute a random subsample of the parent CODEX sample, and that the spectroscopic selection function introduces no significant bias. It also suggests that the dynamical equilibrium state assumption involved in the dynamical mass determination is appropriate on average.

Furthermore, we compute the mean mass ratio ($\beta; M_\text{dyn} / M_\text{wl}$) between the dynamical masses ($M_\text{dyn}$) and the weak lensing masses ($M_\text{wl}$) \citep[e.g.][]{smith2016}, using the results from the scaling relation performed by \citet{capasso2019}.
We obtain a mean mass ratio $\beta = 0.99 \pm 0.03 $ in statistical agreement with $\beta_{X} = 0.95 \pm 0.05$ obtained for the Local Cluster Substructure Survey (LoCuSS) at the redshift $0.15 < z < 0.3$ by \citet{smith2016}.

\section{Conclusion}\label{sec:conclusion}
In this study, we perform a weak lensing analysis of CODEX clusters using the DECaLS DR3 data.
Our cluster sample consists of 279 clusters in the optical richness range 20 $\leq$ $\lambda$ < 110 
and in the redshift range 0.1 $\leq$ $z$ $\leq$ 0.2.
The stacked weak lensing profile of CODEX clusters is computed
in the radial ranges; $1.0 \leq R < 5.5 $ Mpc and $1.0 \leq R < 30.0 $ Mpc,
to avoid the mis-centering effect at smaller scales.

We split our CODEX cluster sample into three optical richness bins: 20 - 30, 30 - 50 and 50 - 110.
In each bin, we measure the stacked excess surface mass density profile ($\Delta \Sigma(R)$). Lacking photometric redshift information for each DECaLS source, we compute an effective critical density based on their redshift distribution. We determine this latter by matching sources to the COSMOS 2015 catalog.
We model the excess surface mass density profile with a NFW profile. 
To reduce the number of free parameters in the fitting, we assume that the concentration-mass relation follows the prescription by \citet{dutton2014}.
We use a MCMC algorithm to estimate the mean cluster masses in each optical richness bin as show in Table \ref{tab:table_cluster}.

We perform a complementary analysis to assess the robustness of our mass estimates. We assume a power-law relation between the mean cluster mass and optical richness $\left\langle M_{\rm 200c} | \lambda \right\rangle = M_{0} \, (\lambda / 40)^{F_{\lambda}}$. 
By fitting an individual cluster, we adjust the free parameters involved in this relation
and restrict the measurement to the inner
region ($1.0 \leq R < 5.5 $ Mpc), and obtain the mean cluster mass at
the pivot richness $\lambda$ = 40,
\begin{eqnarray}
    M_{0} = 3.24_{-0.27}^{+0.29} 
    \times 10^{14} \, \text{M}_\odot \, ,
\end{eqnarray}
and the richness scaling index,
\begin{eqnarray}
    F_{\lambda}= 1.00_{-0.22}^{+0.22} \, .
\end{eqnarray}
We find a good agreement between this scaling relation, and the 
cluster masses determined in the three richness bins. This gives us confidence that our power-law model for the scaling relation is appropriate.

In addition, we compare our scaling relation with the one obtained by \citet{capasso2019} based on a dynamical analysis.
Both relations are in good statistical agreement, although the cluster samples are slightly different. This supports the claim that the SPIDERS subsample used in the dynamical analysis is a random subsample of the parent CODEX catalog used in the weak-lensing analysis.
It also suggests that the dynamical equilibrium assumption involved in the dynamical analysis is appropriate on average.

We also compare the scaling relation of the CODEX clusters 
with the redMaPPer clusters measured by \citet{mcclintock2019}.
The scaling relation for CODEX clusters is shallower than the redMaPPer clusters, which highlights the different selection functions between the two samples. At similar low optical richness, the CODEX clusters are more massive than the redMaPPer clusters.

Our results demonstrate that weak lensing analysis can be performed 
with the DECaLS data. We plan on updating the weak-lensing catalog 
with wider areas and deeper imaging data provided by the final release of 
the DECaLS catalog. Nonetheless, during this analysis, we found the low density of weak-lensing sources in the DEV category given the imaging depth. We would need to extend our calibration to the other categories to increase the source density. Another option would be to run a weak-lensing dedicated shape measurement tool. 

Building on a better weak-lensing catalog, and the advent of forthcoming massive spectroscopic surveys on the same footprint, it becomes tempting to compare weak-lensing and dynamics to high precision in order to study galaxy cluster physics, and test modified gravity models \citep[e.g.][]{pizzuti2019}.

\section*{Acknowledgements}
The authors would like to thank the referees for their valuable comments and suggestions which helped to improve the manuscript, Mauro Sereno and Giovanni Covone for useful and extensive discussions about the weak lensing analysis, and also Nicolas Martinet and Christophe Adami for suggestions that greatly improved this research.

This work is based on the data from the DECaLS DR3 with the Cerro Tololo Inter-American Observatory in Chile under the programs ID: 2013A-0741 and 2014B-0404 and produced by the DECaLS production teams, PI: David Schlegel and Arjun Dey. The Obiwan simulation for the DECaLS data was produced by Kaylan Burleigh.
The CODEX cluster catalog used in this work is based on the RASS and updated by Nathalia Cibirka and the team. The dynamical analysis of the CODEX clusters used in this work produced with the SDSS-IV/SPIDERS program by the SPIDER team. The authors also thank the SPIDERS team for many helpful suggestions on the CODEX cluster catalog.

This work is partially supported by 
Thai Government's Development and Promotion of Science and
Technology Talents Project (DPST),
the National Astronomical Research Institute of Thailand (NARIT)
and the Franco-Thai Scholarship program (French Government grants from the French Embassy in Thailand).
The authors thank the support of the OCEVU Labex (Grant N$^{\circ}$ ANR-11-LABX-0060) and the A*MIDEX project (Grant N$^{\circ}$ ANR-11-IDEX-0001-02) funded by the Investissements d'Avenir French government program managed by the ANR. We also acknowledge support from the ANR eBOSS project (ANR-16-CE31-0021) of the French National Research Agency.

The numerical analysis presented in this work used computing resources at Laboratoire d'Astrophysique de Marseille (LAM) and Centre de Physique des Particules de Marseille (CPPM). Figure \ref{fig:mcmc} was created with ``Corner plot" \citep{corner}.

%%%%%%%%%%%%%%%%%%%%%%%%%%%%%%%%%%%%%%%%%%%%%%%%%%

%%%%%%%%%%%%%%%%%%%% REFERENCES %%%%%%%%%%%%%%%%%%

% The best way to enter references is to use BibTeX:

%\bibliographystyle{mnras}
%\bibliography{example} % if your bibtex file is called example.bib

% Alternatively you could enter them by hand, like this:
% This method is tedious and prone to error if you have lots of references

%%%%%%%%%%%%%%%%%%%%%%%%%%%%%%%%%%%%%%%%%%%%%%%%%%

%%%%%%%%%%%%%%%%% APPENDICES %%%%%%%%%%%%%%%%%%%%%

\appendix

\appendix

%\section{Some extra material}
%If you want to present additional material which would interrupt the flow of the main paper,
%it can be placed in an Appendix which appears after the list of references.
%

%
%
\section{Ellipticity correction}\label{sec:correction_factor}
\begin{figure}
\centering
	\includegraphics[width=\columnwidth]{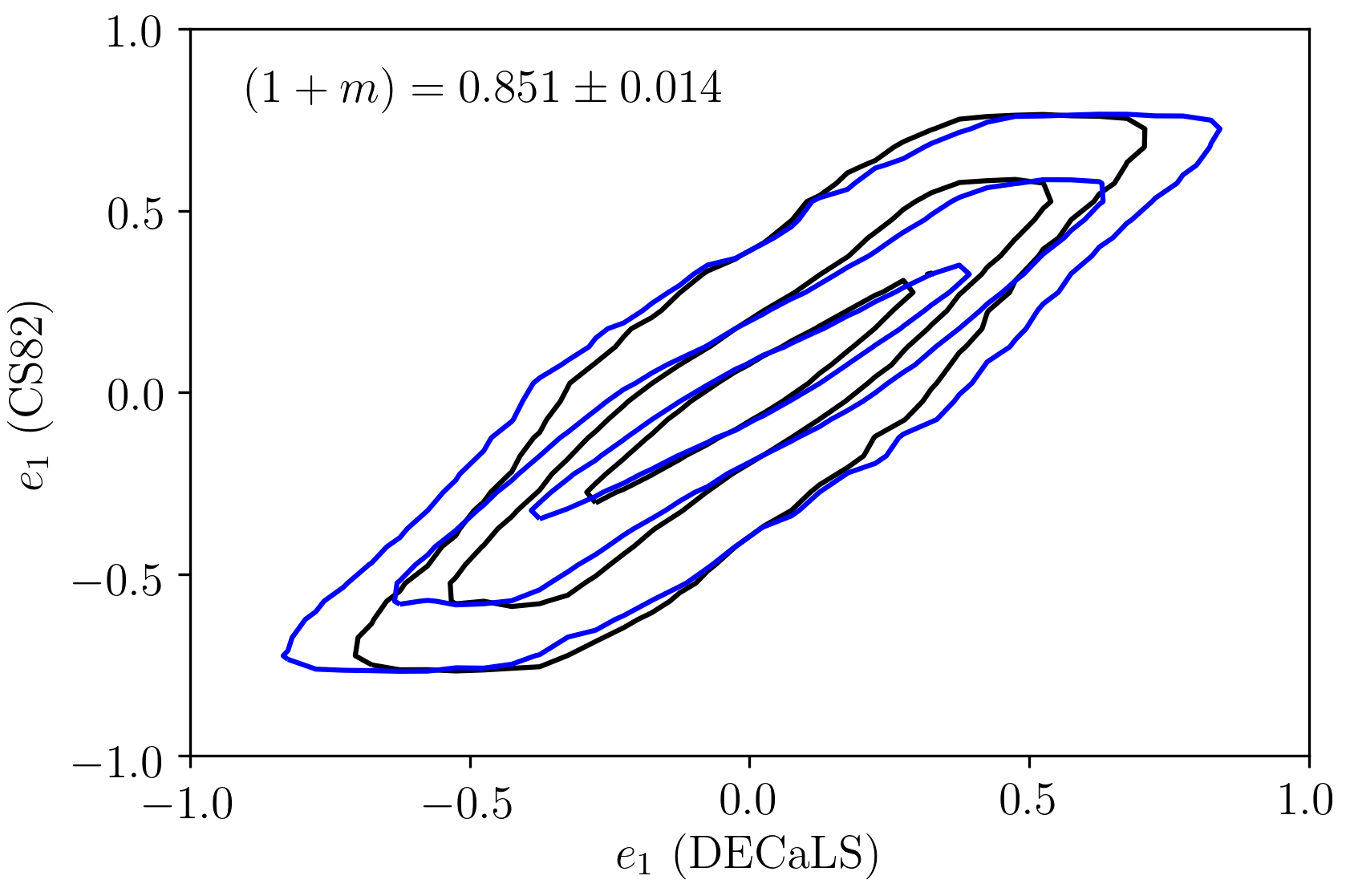}
    \caption{Comparisons of the ellipticity 1 ($e_{1}$) for the EXP object in the DECaLS DR3 shear catalog before (black solid lines) and after correction (blue solid lines) with the CS82 data. The contour plots present the confidence levels at $1\sigma$, $2\sigma$ and $3\sigma$ respectively.}
    \label{fig:e1_calibration}
\end{figure}

The ellipticity parameter was measured for each source. 
It is different from the usual ellipticity definition in the DECaLS DR3 catalog which is given by
\begin{eqnarray}
	\varepsilon \equiv \sqrt{1 - (b/a)^{2}} \, ,
\end{eqnarray}
where $a$ and $b$ are the semi-major and -minor axis of an elliptical source (galaxy).
For a gravitational analysis, the ellipticity is a complex number,
\begin{eqnarray}
\varepsilon = \frac{a - b}{a + b} \text{exp} (2i\phi) = \varepsilon_{1} + i\varepsilon_{2} \, ,
\end{eqnarray}
where $\phi$ is the position angle relative to the reference frame with a range of $0^{\circ}$ to $180^{\circ}$.
This ellipticity has been used in the weak lensing analysis.
However, we must calibrate it before using in the measurement.
We calibrated the DECaLS DR3 sources with the CS82 data.
We define the relation between the observed and corrected ellipticities as
\begin{eqnarray}
	\varepsilon_{1, \text{corr}} = (1 + m) \varepsilon_{1, \text{obs}},\\
	\varepsilon_{2, \text{corr}} = (1 + m) \varepsilon_{2, \text{obs}} + c_{2},
\end{eqnarray}
where $(1 + m)$ is defined in Equation \ref{eq:oneplusm} and the $c_{2}$ component is defined as
\begin{eqnarray}
	c_{2} = b_{0} + (b_{1} \times magz) + (b_{2} \times magz^{2}) \, .
\end{eqnarray}
The calibration parameters of the DECaLS DR3 catalog with the CS82 catalog
are shown in Table \ref{tab:ellipticity_decals} and the $c_{2}$ component is given by
$b_{0}$ = 0.25577, $b_{1}$ = -0.02266, $b_{2}$ = 0.00050 with the magnitude cut $magz$ < 21.0. In Figure \ref{fig:e1_calibration}, we compare the ellipticity 1 ($e_{1}$) between the CS82 and DECaLS catalog before (black solid lines) and after correction (blue solid lines) with the CS82 data for the EXP object in the DECaLS DR3 shear catalog.
\begin{table*}
\caption{The calibration parameters of the DECaLS DR3 shear catalog with the CS82 catalog and Obiwan simulation.}
\label{tab:ellipticity_decals}
\begin{tabular}{cccccccc}
\hline
Type & Number of objects & $a_{0}$ (CS82)& $a_{1}$ (CS82) & 1 + m (CS82) & 1 + m (Obiwan)
& $c_{2}$ (CS82) & $c_{2}$ (Obiwan)\\
\hline
SIMP  & 25462647 & 1.23907 & 0.02817  & 0.697 $\pm$ 0.099 & - & 0.002 $\pm$ 0.002 & -\\
EXP   & 24923051 & 1.12193 & -0.00011 & 0.851 $\pm$ 0.014 & 0.896 $\pm$ 0.045 & 0.002 $\pm$ 0.002 & 0.000 $\pm$ 0.001 \\
DEV   & 5404940 & 1.18554 & 0.00946   & 0.771 $\pm$ 0.078 & - & 0.004 $\pm$ 0.003 & -\\
COMP  & 75429  & 1.18554 & 0.00946    & 0.773 $\pm$ 0.108 & - & 0.007 $\pm$ 0.004 & -\\
\hline
\end{tabular}
\end{table*}
In addition, the data from DECaLS DR3 catalog were tested 
with the Obiwan simulations\footnote{\url{https://obiwan.readthedocs.io}}.
This simulation simulated ELG galaxies with Sersic profiles on top of real DECam images, and re-run the Tractor tool to produce catalogs. After matching the simulated positions and the Tractor positions 
within a 5 arcsec radius, we obtained approximately 100,000 galaxies for the calibration.
From the fitting, we obtained best fit values for the EXP objects, $a_{0}$ = 1.33320, $a_{1}$ = 0.00656. 
For the $c_{2}$ component, we obtained $b_{0}$ = 0.31341, $b_{1}$ = -0.02914, $b_{2}$ = 0.00067, 
with the cut in magnitude $magz$ < 21.3.

\section{Miscentering effect}
\label{sec:miscentering}
By stacking the weak lensing profile, the cluster center should be correctly defined to provide the true mass profile.
In this work, we choose the brightest cluster galaxy (BCG) as the cluster center.
However, the center of mass distribution in galaxy clusters can be shifted from the BCG, due to physical processes in cluster cores. 
In this section, we assess the impact of mis-centering, and demonstrate that our radial cut at $R>1$ Mpc, safely mitigate 
any bias due to this effect.
Note that the redMaPPer algorithm computed a probability in each galaxy to be the BCG and selects the most likely as the BCG for that cluster.
However, the miscentering rate of the redMaPPer galaxy clusters is about $f_{\rm mis} = 0.25$ (\citealp{rykoff2014, rozo2014}).

To model the miscentering effect on the cluster mass distribution, we adopt the expression given by
\begin{eqnarray}
\begin{split}
\label{eq:full_model}
	\Delta \Sigma(R) =  (1 - f_{\text{mis}})\Delta \Sigma_{\text{1h}}^{\text{NFW}}(R) 
	+ f_{\text{mis}}\Delta \Sigma_{\text{mis}}^{\text{NFW}}(R) \\
	+ \Delta \Sigma_{\text{2h}}(R) \, ,
\end{split}
\end{eqnarray}
where $\Delta \Sigma_{\text{1h}}^{\text{NFW}}$ 
is the excess surface mass density of NFW profile,
$\Delta \Sigma_{\text{mis}}^{\text{NFW}}$ is the miscentering profile, $\Delta \Sigma_{\text{2h}}$ is the excess surface mass profile of the second dark matter halo and $f_{\text{mis}}$ is the miscentering factor.
Formally, the second halo term should also be miscentered, 
but the effect is so weak that we decided to only apply this effect to the first halo term.

\begin{figure}
\includegraphics[width=\columnwidth]{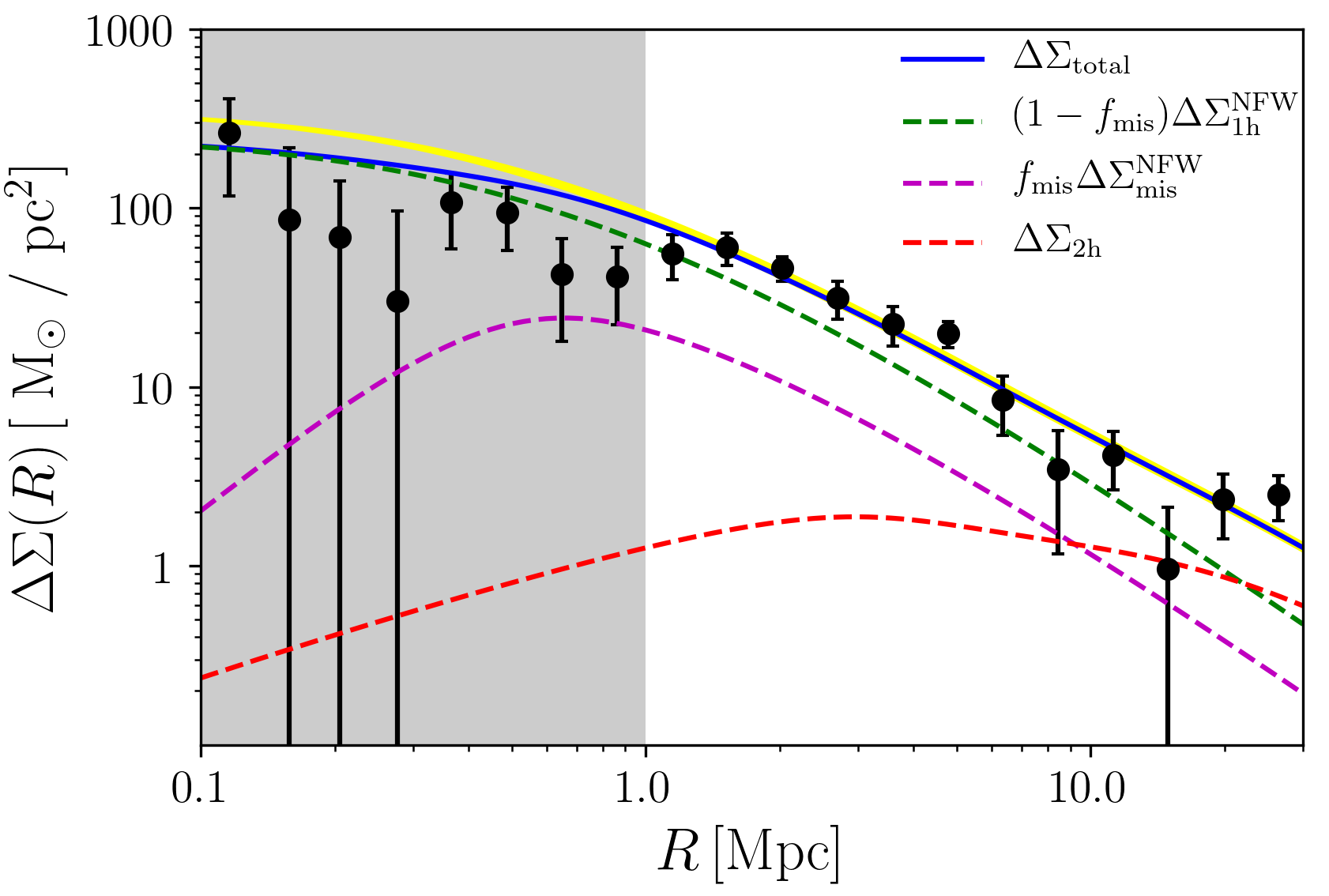}
\caption{The excess surface mass density profile from our CODEX cluster sample for the optical richness bin $\lambda = 50 - 110$ (black dots). The blue solid line shows the theoretical profile defined in Equation \ref{eq:full_model}, by assuming the miscentering factor $f_{\text{mis}} = 0.25$, $\sigma_{\text{mis}} = 0.21$ and $M_{\rm 200c} = 5.02 \times 10^{14} \text{M}_{\odot}$. The dotted lines show each term in following equation contributing to the profile. The yellow line shows the theoretical profile defined by two terms as in Equation \ref{eq:sigma_stacking} with the same cluster mass. The grey shaded area was excluded from our analysis due to the lack of the sources and high contamination in that region.}
\label{fig:full_test_model}
\end{figure}

The miscentering profile is given by the projected excess surface mass density of the NFW profile which can be written by
\begin{eqnarray}
    \Delta \Sigma_{\text{mis}}^{\text{NFW}}(R | P(R_{\text{mis}})) = \int_{0}^{\infty} \, \Sigma(R | R_{\text{mis}}) \, P(R_{\text{mis}}) \, \text{d}R_{\text{mis}} \, ,
\end{eqnarray}
where $P(R_{\text{mis}})$ is the miscentering distribution chosen by the Rayleigh distribution function with the miscentering radius ($R_{\text{mis}}$) and parameter ($\sigma_{\text{mis}}$) as
\begin{eqnarray}
    P(R_{\text{mis}}) =\frac{R_{\text{mis}}}{\sigma_{\text{mis}}^{2}} \exp 
    \left[ -\frac{1}{2} \left(\frac{R_{\text{mis}}}{\sigma_{\text{mis}}}\right)^{2} \right] \, .
\end{eqnarray}
Accordingly, the excess surface mass density of miscentering term of the NFW profile is
\begin{eqnarray}
    \Sigma(R | R_{\text{mis}}) = \frac{1}{2\pi} \int_{0}^{2\pi}
    \Sigma(r) \, \text{d} \theta \, ,
\end{eqnarray}
where $r = \sqrt{R^2 + R_{\text{mis}}^2 - 2RR_{\text{mis}}\cos(\theta)}$ is the projected radius at the coordinates ($R, \theta$) related to the miscentering radius (\citealp[see e.g.][]{yang2006, johnston2007, matthew2012, mcclintock2019}).
In Figure \ref{fig:full_test_model}, we plot the stacked excess surface mass density of the CODEX cluster sample for the optical richness bin $\lambda = 50 - 110$ as in Section \ref{sec:results_samplecodex}.
We assess the miscentering effect with 
$f_{\text{mis}} = 0.25$ and
$\sigma_{\rm mis} = 0.21$ Mpc (\citealp[i.e.][]{mcclintock2019, zhang2019}).
The lack of sources and high contamination in the inner region yield
a very noisy signal at $R<1$ Mpc (grey shaded area). Consequently, we decided to exclude this radial range to avoid any bias in our mass estimates.

In our analysis, we adjust a theoretical profile made of only two terms ($\Delta \Sigma_{\text{1h}}^{\text{NFW}} + \Delta \Sigma_{\text{2h}}$) as shown in Equation \ref{eq:sigma_stacking}. 
It is sufficient to reproduce the weak lensing signal in the radius range $R > 1$ Mpc.
As a result, we decided not to include the miscentering effect.

\section{Systematic Tests}
\subsection{Testing on the non-lensing mode}
\begin{figure}\centering
	\includegraphics[width=\columnwidth]{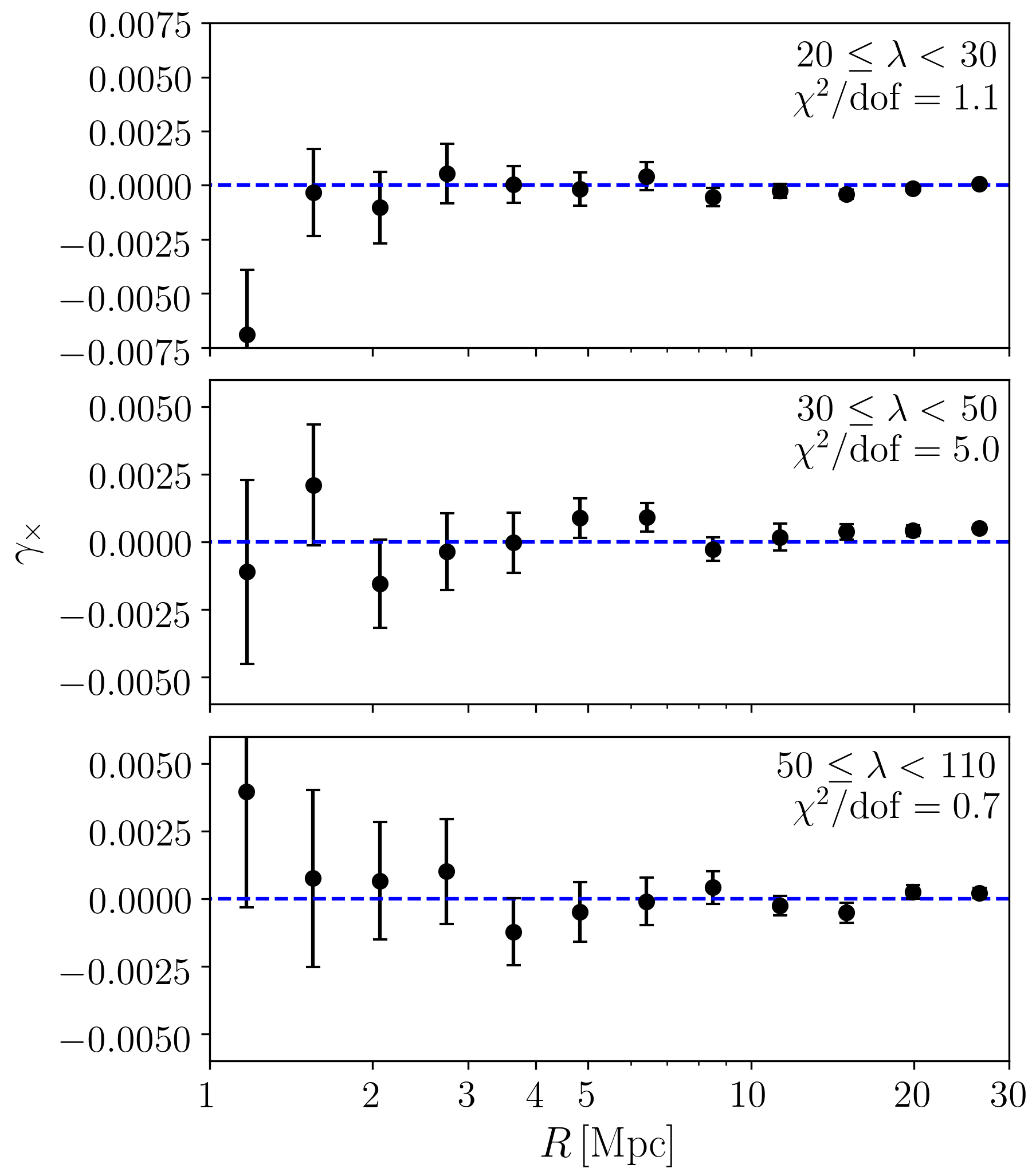}
    \caption{The cross shear profiles per radial bin of the stacked clusters in three richness groups as described in Section \ref{sec:results_samplecodex}. The error bars estimated from the Jackknife covariance matrix and the degree of freedom is equal to 12 for 12 radial bins.}
    \label{fig:cross_shear}
\end{figure}

In this section, we perform a systematic test on the ellipticity parameters of the DECaLS DR3 shear catalog. 
The non-lensing mode of a shear component (cross shear) has been computed 
by rotating the ellipticities 45 degrees.
In theory, gravitational lensing does not produce a cross shear component.
Therefore, we can use the cross shear component to reveal biases in the shape measurement operation.
We compute the tangential shear ($\gamma_{t}$)
and the cross shear ($\gamma_{\times}$) from the real and imaginary parts of the ellipticity measurements 
\begin{eqnarray}
\gamma_{t} = -\textbf{Re}[\gamma e^{-2i\phi}] \,\,\, \text{and} \,\,\, \gamma_{\times} = -\textbf{Im}[\gamma e^{-2i\phi}] \, .
\end{eqnarray}
We split our CODEX cluster sample in richness bins as in Section \ref{sec:results_samplecodex}; $\lambda$ = 20 - 30, 30 - 50 and 50 - 110.
We plot the cross shear component in Figure \ref{fig:cross_shear}. It is statistically consistent with zero at all radii and for our three richness bins.

\subsection{The dilution effect by lens-source galaxies}
\label{sec:dilution_galaxies}
\begin{figure}\centering
	\includegraphics[width=\columnwidth]{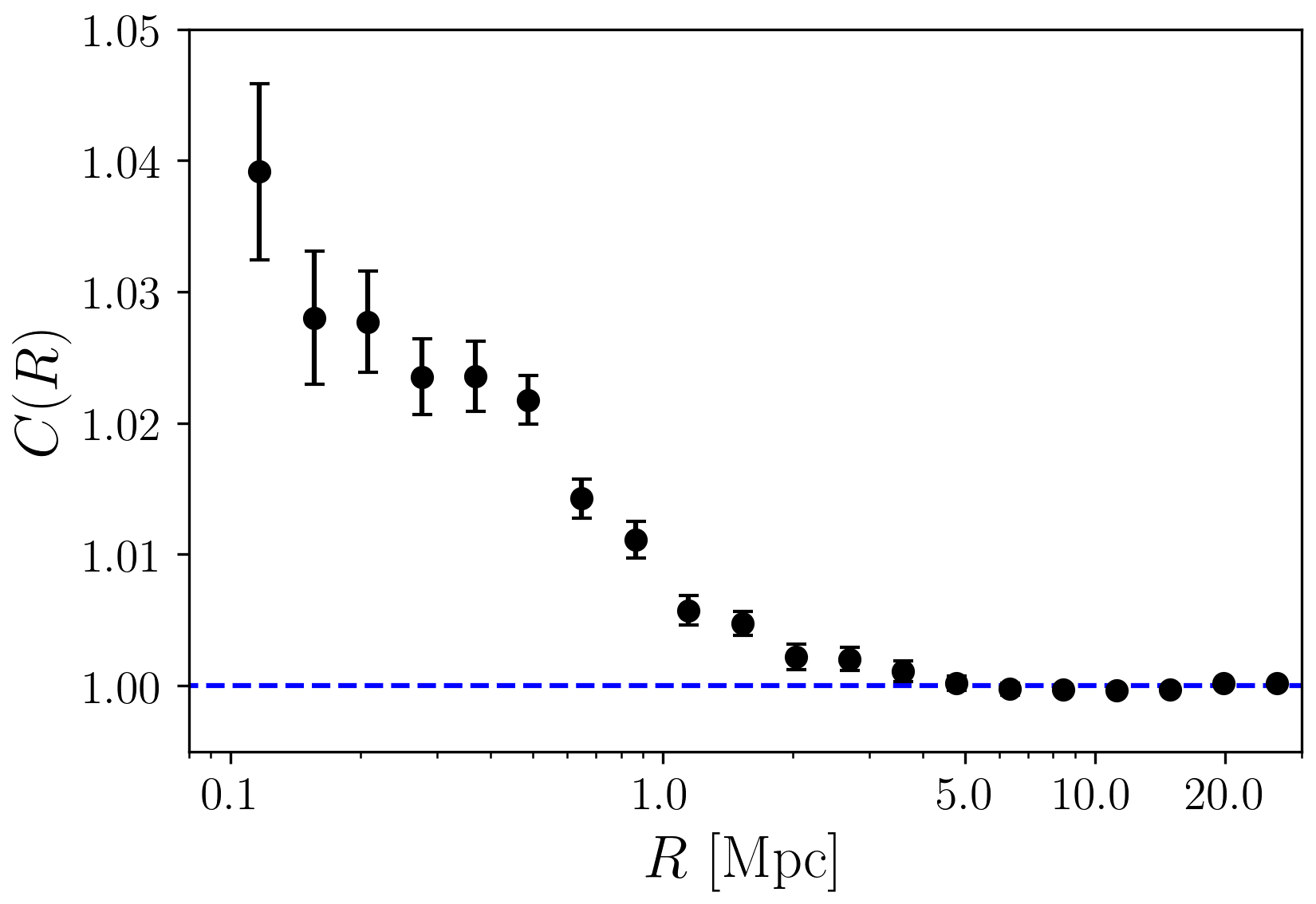}
    \caption{The correction factor of CODEX cluster samples as a function of cluster radius.}
    \label{fig:correction_factor}
\end{figure}

As we discussed in Section ~\ref{sec:wl_estimator}, the correction factor $C(R)$ is used to correct the contamination in the weak-lensing signal produced by the overdensity of galaxies in the cluster center \citep{simet2017}.
This effect dilutes the amplitude of the shear signal, especially,
in the inner region of a galaxy cluster.
In this test, we stack CODEX cluster samples and extend the shear measurement to smaller radius in the range $0.1 \leq R < 30$ Mpc.
The results in Figure \ref{fig:correction_factor} show that in the inner region, the dilution goes up to $\approx$ 4\%.
However, in the outer region (ex. $R >$ 1 Mpc), the correction factor is less than 1\%. 
Our measurements are therefore very little affected by the lens-source dilution effect.

%%%%%%%%%%%%%%%%%%%%%%%%%%%%%%%%%%%%%%%%%%%%%%%%%%

\bibliographystyle{mnras}
\bibliography{references}

\begin{thebibliography}{}
\makeatletter
\relax
\def\mn@urlcharsother{\let\do\@makeother \do\$\do\&\do\#\do\^\do\_\do\%\do\~}
\def\mn@doi{\begingroup\mn@urlcharsother \@ifnextchar [ {\mn@doi@}
  {\mn@doi@[]}}
\def\mn@doi@[#1]#2{\def\@tempa{#1}\ifx\@tempa\@empty \href
  {http://dx.doi.org/#2} {doi:#2}\else \href {http://dx.doi.org/#2} {#1}\fi
  \endgroup}
\def\mn@eprint#1#2{\mn@eprint@#1:#2::\@nil}
\def\mn@eprint@arXiv#1{\href {http://arxiv.org/abs/#1} {{\tt arXiv:#1}}}
\def\mn@eprint@dblp#1{\href {http://dblp.uni-trier.de/rec/bibtex/#1.xml}
  {dblp:#1}}
\def\mn@eprint@#1:#2:#3:#4\@nil{\def\@tempa {#1}\def\@tempb {#2}\def\@tempc
  {#3}\ifx \@tempc \@empty \let \@tempc \@tempb \let \@tempb \@tempa \fi \ifx
  \@tempb \@empty \def\@tempb {arXiv}\fi \@ifundefined
  {mn@eprint@\@tempb}{\@tempb:\@tempc}{\expandafter \expandafter \csname
  mn@eprint@\@tempb\endcsname \expandafter{\@tempc}}}

\bibitem[\protect\citeauthoryear{{Allen}, {Rapetti}, {Schmidt}, {Ebeling},
  {Morris}  \& {Fabian}}{{Allen} et~al.}{2008}]{allen2008}
{Allen} S.~W.,  {Rapetti} D.~A.,  {Schmidt} R.~W.,  {Ebeling} H.,  {Morris}
  R.~G.,   {Fabian} A.~C.,  2008, \mn@doi [\mnras]
  {10.1111/j.1365-2966.2007.12610.x}, \href
  {http://adsabs.harvard.edu/abs/2008MNRAS.383..879A} {383, 879}

\bibitem[\protect\citeauthoryear{{Angulo}, {Springel}, {White}, {Jenkins},
  {Baugh}  \& {Frenk}}{{Angulo} et~al.}{2012}]{angulo2012}
{Angulo} R.~E.,  {Springel} V.,  {White} S.~D.~M.,  {Jenkins} A.,  {Baugh}
  C.~M.,   {Frenk} C.~S.,  2012, \mn@doi [\mnras]
  {10.1111/j.1365-2966.2012.21830.x}, \href
  {https://ui.adsabs.harvard.edu/abs/2012MNRAS.426.2046A} {426, 2046}

\bibitem[\protect\citeauthoryear{{Baxter} et~al.,}{{Baxter}
  et~al.}{2015}]{baxter2015}
{Baxter} E.~J.,  et~al., 2015, \mn@doi [The Astrophysical Journal]
  {10.1088/0004-637X/806/2/247}, \href
  {https://ui.adsabs.harvard.edu/abs/2015ApJ...806..247B} {806, 247}

\bibitem[\protect\citeauthoryear{{Becker} \& {Kravtsov}}{{Becker} \&
  {Kravtsov}}{2011}]{becker2011}
{Becker} M.~R.,  {Kravtsov} A.~V.,  2011, \mn@doi [The Astrophysical Journal]
  {10.1088/0004-637X/740/1/25}, \href
  {https://ui.adsabs.harvard.edu/abs/2011ApJ...740...25B} {740, 25}

\bibitem[\protect\citeauthoryear{{Bertin}}{{Bertin}}{2011}]{bertin2011}
{Bertin} E.,  2011, in {Evans} I.~N.,  {Accomazzi} A.,  {Mink} D.~J.,   {Rots}
  A.~H.,  eds,  Astronomical Society of the Pacific Conference Series Vol. 442,
  Astronomical Data Analysis Software and Systems XX. p.~435

\bibitem[\protect\citeauthoryear{{Bhattacharya}, {Heitmann}, {White},
  {Luki{\'c}}, {Wagner}  \& {Habib}}{{Bhattacharya}
  et~al.}{2011}]{bhattacharya2011}
{Bhattacharya} S.,  {Heitmann} K.,  {White} M.,  {Luki{\'c}} Z.,  {Wagner} C.,
   {Habib} S.,  2011, \mn@doi [\apj] {10.1088/0004-637X/732/2/122}, \href
  {https://ui.adsabs.harvard.edu/abs/2011ApJ...732..122B} {732, 122}

\bibitem[\protect\citeauthoryear{{Biviano} et~al.,}{{Biviano}
  et~al.}{2013}]{biviano2013}
{Biviano} A.,  et~al., 2013, \mn@doi [\aap] {10.1051/0004-6361/201321955},
  \href {https://ui.adsabs.harvard.edu/abs/2013A&A...558A...1B} {558, A1}

\bibitem[\protect\citeauthoryear{{Blanton} et~al.,}{{Blanton}
  et~al.}{2017}]{blanton2017}
{Blanton} M.~R.,  et~al., 2017, \mn@doi [\aj] {10.3847/1538-3881/aa7567}, \href
  {https://ui.adsabs.harvard.edu/abs/2017AJ....154...28B} {154, 28}

\bibitem[\protect\citeauthoryear{{Bose} et~al.,}{{Bose}
  et~al.}{2017}]{bose2017}
{Bose} S.,  et~al., 2017, \mn@doi [\mnras] {10.1093/mnras/stw2686}, \href
  {https://ui.adsabs.harvard.edu/\#abs/2017MNRAS.464.4520B} {464, 4520}

\bibitem[\protect\citeauthoryear{{Capasso} et~al.,}{{Capasso}
  et~al.}{2019}]{capasso2019}
{Capasso} R.,  et~al., 2019, \mn@doi [\mnras] {10.1093/mnras/stz931}, \href
  {https://ui.adsabs.harvard.edu/abs/2019MNRAS.tmp..901C} {p.~901}

\bibitem[\protect\citeauthoryear{{Cibirka} et~al.,}{{Cibirka}
  et~al.}{2017}]{cibirka2017}
{Cibirka} N.,  et~al., 2017, \mn@doi [\mnras] {10.1093/mnras/stx484}, \href
  {http://adsabs.harvard.edu/abs/2017MNRAS.468.1092C} {468, 1092}

\bibitem[\protect\citeauthoryear{{Clerc} et~al.,}{{Clerc}
  et~al.}{2016}]{clerc2016}
{Clerc} N.,  et~al., 2016, \mn@doi [\mnras] {10.1093/mnras/stw2214}, \href
  {https://ui.adsabs.harvard.edu/\#abs/2016MNRAS.463.4490C} {463, 4490}

\bibitem[\protect\citeauthoryear{{Dawson} et~al.,}{{Dawson}
  et~al.}{2016}]{dawson2016}
{Dawson} K.~S.,  et~al., 2016, \mn@doi [\aj] {10.3847/0004-6256/151/2/44},
  \href {https://ui.adsabs.harvard.edu/abs/2016AJ....151...44D} {151, 44}

\bibitem[\protect\citeauthoryear{{Dekel} \& {Lahav}}{{Dekel} \&
  {Lahav}}{1999}]{dekel1999}
{Dekel} A.,  {Lahav} O.,  1999, \mn@doi [\apj] {10.1086/307428}, \href
  {https://ui.adsabs.harvard.edu/\#abs/1999ApJ...520...24D} {520, 24}

\bibitem[\protect\citeauthoryear{{Despali}, {Giocoli}, {Angulo}, {Tormen},
  {Sheth}, {Baso}  \& {Moscardini}}{{Despali} et~al.}{2016}]{despali2016}
{Despali} G.,  {Giocoli} C.,  {Angulo} R.~E.,  {Tormen} G.,  {Sheth} R.~K.,
  {Baso} G.,   {Moscardini} L.,  2016, \mn@doi [\mnras]
  {10.1093/mnras/stv2842}, \href
  {https://ui.adsabs.harvard.edu/\#abs/2016MNRAS.456.2486D} {456, 2486}

\bibitem[\protect\citeauthoryear{{Dey} et~al.,}{{Dey} et~al.}{2019}]{dey2019}
{Dey} A.,  et~al., 2019, \mn@doi [\aj] {10.3847/1538-3881/ab089d}, \href
  {http://adsabs.harvard.edu/abs/2019AJ....157..168D} {157, 168}

\bibitem[\protect\citeauthoryear{{Diemer} \& {Kravtsov}}{{Diemer} \&
  {Kravtsov}}{2014}]{diemer2014}
{Diemer} B.,  {Kravtsov} A.~V.,  2014, \mn@doi [The Astrophysical Journal]
  {10.1088/0004-637X/789/1/1}, \href
  {https://ui.adsabs.harvard.edu/abs/2014ApJ...789....1D} {789, 1}

\bibitem[\protect\citeauthoryear{{Donahue} et~al.,}{{Donahue}
  et~al.}{2014}]{donahue2014}
{Donahue} M.,  et~al., 2014, \mn@doi [\apj] {10.1088/0004-637X/794/2/136},
  \href {https://ui.adsabs.harvard.edu/abs/2014ApJ...794..136D} {794, 136}

\bibitem[\protect\citeauthoryear{{Duffy}, {Schaye}, {Kay}  \& {Dalla
  Vecchia}}{{Duffy} et~al.}{2008}]{duffy2008}
{Duffy} A.~R.,  {Schaye} J.,  {Kay} S.~T.,   {Dalla Vecchia} C.,  2008, \mn@doi
  [\mnras] {10.1111/j.1745-3933.2008.00537.x}, \href
  {https://ui.adsabs.harvard.edu/#abs/2008MNRAS.390L..64D} {390, L64}

\bibitem[\protect\citeauthoryear{{Dutton} \& {Macci{\`o}}}{{Dutton} \&
  {Macci{\`o}}}{2014}]{dutton2014}
{Dutton} A.~A.,  {Macci{\`o}} A.~V.,  2014, \mn@doi [\mnras]
  {10.1093/mnras/stu742}, \href
  {http://adsabs.harvard.edu/abs/2014MNRAS.441.3359D} {441, 3359}

\bibitem[\protect\citeauthoryear{{Einasto}}{{Einasto}}{1965}]{einasto1965}
{Einasto} J.,  1965, Trudy Astrofizicheskogo Instituta Alma-Ata, \href
  {https://ui.adsabs.harvard.edu/abs/1965TrAlm...5...87E} {5, 87}

\bibitem[\protect\citeauthoryear{{Euclid Collaboration} et~al.,}{{Euclid
  Collaboration} et~al.}{2019}]{martinet2019}
{Euclid Collaboration} et~al., 2019, arXiv e-prints, \href
  {https://ui.adsabs.harvard.edu/abs/2019arXiv190200044E} {p. arXiv:1902.00044}

\bibitem[\protect\citeauthoryear{{Falco}, {Mamon}, {Wojtak}, {Hansen}  \&
  {Gottl{\"o}ber}}{{Falco} et~al.}{2013}]{falco2013}
{Falco} M.,  {Mamon} G.~A.,  {Wojtak} R.,  {Hansen} S.~H.,   {Gottl{\"o}ber}
  S.,  2013, \mn@doi [\mnras] {10.1093/mnras/stt1768}, \href
  {https://ui.adsabs.harvard.edu/abs/2013MNRAS.436.2639F} {436, 2639}

\bibitem[\protect\citeauthoryear{{Farahi}, {Evrard}, {Rozo}, {Rykoff}  \&
  {Wechsler}}{{Farahi} et~al.}{2016}]{farahi2016}
{Farahi} A.,  {Evrard} A.~E.,  {Rozo} E.,  {Rykoff} E.~S.,   {Wechsler} R.~H.,
  2016, \mn@doi [Monthly Notices of the Royal Astronomical Society]
  {10.1093/mnras/stw1143}, \href
  {https://ui.adsabs.harvard.edu/abs/2016MNRAS.460.3900F} {460, 3900}

\bibitem[\protect\citeauthoryear{{Fischer} et~al.,}{{Fischer}
  et~al.}{2000}]{fischer2000}
{Fischer} P.,  et~al., 2000, \mn@doi [\aj] {10.1086/301540}, \href
  {http://adsabs.harvard.edu/abs/2000AJ....120.1198F} {120, 1198}

\bibitem[\protect\citeauthoryear{Foreman-Mackey}{Foreman-Mackey}{2016}]{corner}
Foreman-Mackey D.,  2016, \mn@doi [The Journal of Open Source Software]
  {10.21105/joss.00024}, 24

\bibitem[\protect\citeauthoryear{{Foreman-Mackey}, {Hogg}, {Lang}  \&
  {Goodman}}{{Foreman-Mackey} et~al.}{2013}]{daniel2013}
{Foreman-Mackey} D.,  {Hogg} D.~W.,  {Lang} D.,   {Goodman} J.,  2013, \mn@doi
  [Publications of the Astronomical Society of the Pacific] {10.1086/670067},
  \href {https://ui.adsabs.harvard.edu/\#abs/2013PASP..125..306F} {125, 306}

\bibitem[\protect\citeauthoryear{{Geach} \& {Peacock}}{{Geach} \&
  {Peacock}}{2017}]{geach2017}
{Geach} J.~E.,  {Peacock} J.~A.,  2017, \mn@doi [Nature Astronomy]
  {10.1038/s41550-017-0259-1}, \href
  {https://ui.adsabs.harvard.edu/abs/2017NatAs...1..795G} {1, 795}

\bibitem[\protect\citeauthoryear{{George} et~al.,}{{George}
  et~al.}{2012}]{matthew2012}
{George} M.~R.,  et~al., 2012, \mn@doi [\apj] {10.1088/0004-637X/757/1/2},
  \href {https://ui.adsabs.harvard.edu/abs/2012ApJ...757....2G} {757, 2}

\bibitem[\protect\citeauthoryear{{Hartlap}, {Simon}  \& {Schneider}}{{Hartlap}
  et~al.}{2007}]{hartlap2007}
{Hartlap} J.,  {Simon} P.,   {Schneider} P.,  2007, \mn@doi [Astronomy and
  Astrophysics] {10.1051/0004-6361:20066170}, \href
  {https://ui.adsabs.harvard.edu/abs/2007A&A...464..399H} {464, 399}

\bibitem[\protect\citeauthoryear{{Hayashi} \& {White}}{{Hayashi} \&
  {White}}{2008}]{hayashi2008}
{Hayashi} E.,  {White} S.~D.~M.,  2008, \mn@doi [\mnras]
  {10.1111/j.1365-2966.2008.13371.x}, \href
  {https://ui.adsabs.harvard.edu/abs/2008MNRAS.388....2H} {388, 2}

\bibitem[\protect\citeauthoryear{{Heymans} et~al.,}{{Heymans}
  et~al.}{2012}]{heymans2012}
{Heymans} C.,  et~al., 2012, \mn@doi [\mnras]
  {10.1111/j.1365-2966.2012.21952.x}, \href
  {https://ui.adsabs.harvard.edu/\#abs/2012MNRAS.427..146H} {427, 146}

\bibitem[\protect\citeauthoryear{{Hoekstra}, {Hartlap}, {Hilbert}  \& {van
  Uitert}}{{Hoekstra} et~al.}{2011}]{hoekstra2011}
{Hoekstra} H.,  {Hartlap} J.,  {Hilbert} S.,   {van Uitert} E.,  2011, \mn@doi
  [\mnras] {10.1111/j.1365-2966.2010.18053.x}, \href
  {http://adsabs.harvard.edu/abs/2011MNRAS.412.2095H} {412, 2095}

\bibitem[\protect\citeauthoryear{{Ilbert} et~al.,}{{Ilbert}
  et~al.}{2013}]{ilbert2013}
{Ilbert} O.,  et~al., 2013, \mn@doi [\aap] {10.1051/0004-6361/201321100}, \href
  {https://ui.adsabs.harvard.edu/abs/2013A%26A...556A..55I} {556, A55}

\bibitem[\protect\citeauthoryear{{Jauzac} et~al.,}{{Jauzac}
  et~al.}{2012}]{jauzac2012}
{Jauzac} M.,  et~al., 2012, \mn@doi [\mnras]
  {10.1111/j.1365-2966.2012.21966.x}, \href
  {https://ui.adsabs.harvard.edu/\#abs/2012MNRAS.426.3369J} {426, 3369}

\bibitem[\protect\citeauthoryear{{Johnston} et~al.,}{{Johnston}
  et~al.}{2007}]{johnston2007}
{Johnston} D.~E.,  et~al., 2007, arXiv e-prints, \href
  {https://ui.adsabs.harvard.edu/\#abs/2007arXiv0709.1159J} {p.
  arXiv:0709.1159}

\bibitem[\protect\citeauthoryear{{Jullo} et~al.,}{{Jullo}
  et~al.}{2012}]{jullo2012}
{Jullo} E.,  et~al., 2012, \mn@doi [\apj] {10.1088/0004-637X/750/1/37}, \href
  {https://ui.adsabs.harvard.edu/\#abs/2012ApJ...750...37J} {750, 37}

\bibitem[\protect\citeauthoryear{{Klein}, {Grandis}, {Mohr}, {Paulus}  \& {the
  DES Collaboration}}{{Klein} et~al.}{2018}]{klein2018}
{Klein} M.,  {Grandis} S.,  {Mohr} J.,  {Paulus} M.,   {the DES Collaboration}
  2018, arXiv e-prints, \href
  {https://ui.adsabs.harvard.edu/abs/2018arXiv181209956K} {p. arXiv:1812.09956}

\bibitem[\protect\citeauthoryear{{Laigle} et~al.,}{{Laigle}
  et~al.}{2016}]{laigle2016}
{Laigle} C.,  et~al., 2016, \mn@doi [The Astrophysical Journal Supplement
  Series] {10.3847/0067-0049/224/2/24}, \href
  {https://ui.adsabs.harvard.edu/\#abs/2016ApJS..224...24L} {224, 24}

\bibitem[\protect\citeauthoryear{{Lang}, {Hogg}  \& {Schlegel}}{{Lang}
  et~al.}{2014}]{lang2014}
{Lang} D.,  {Hogg} D.~W.,   {Schlegel} D.~J.,  2014, arXiv e-prints, \href
  {https://ui.adsabs.harvard.edu/abs/2014arXiv1410.7397L} {p. arXiv:1410.7397}

\bibitem[\protect\citeauthoryear{{Lewis} \& {Challinor}}{{Lewis} \&
  {Challinor}}{2011}]{camb2011}
{Lewis} A.,  {Challinor} A.,  2011, {CAMB: Code for Anisotropies in the
  Microwave Background}, Astrophysics Source Code Library (\mn@eprint {ascl}
  {1102.026})

\bibitem[\protect\citeauthoryear{{Limousin}, {Morandi}, {Sereno}, {Meneghetti},
  {Ettori}, {Bartelmann}  \& {Verdugo}}{{Limousin} et~al.}{2013}]{limousin2013}
{Limousin} M.,  {Morandi} A.,  {Sereno} M.,  {Meneghetti} M.,  {Ettori} S.,
  {Bartelmann} M.,   {Verdugo} T.,  2013, \mn@doi [\ssr]
  {10.1007/s11214-013-9980-y}, \href
  {https://ui.adsabs.harvard.edu/\#abs/2013SSRv..177..155L} {177, 155}

\bibitem[\protect\citeauthoryear{{Liu} et~al.,}{{Liu} et~al.}{2015}]{liu2015}
{Liu} X.,  et~al., 2015, \mn@doi [\mnras] {10.1093/mnras/stv784}, \href
  {https://ui.adsabs.harvard.edu/\#abs/2015MNRAS.450.2888L} {450, 2888}

\bibitem[\protect\citeauthoryear{{Mamon}, {Biviano}  \& {Bou{\'e}}}{{Mamon}
  et~al.}{2013}]{mamon2013}
{Mamon} G.~A.,  {Biviano} A.,   {Bou{\'e}} G.,  2013, \mn@doi [\mnras]
  {10.1093/mnras/sts565}, \href
  {https://ui.adsabs.harvard.edu/abs/2013MNRAS.429.3079M} {429, 3079}

\bibitem[\protect\citeauthoryear{{Mandelbaum} et~al.,}{{Mandelbaum}
  et~al.}{2005}]{mandelbaum2005}
{Mandelbaum} R.,  et~al., 2005, \mn@doi [\mnras]
  {10.1111/j.1365-2966.2005.09282.x}, \href
  {http://adsabs.harvard.edu/abs/2005MNRAS.361.1287M} {361, 1287}

\bibitem[\protect\citeauthoryear{{Mandelbaum}, {Seljak}, {Cool}, {Blanton},
  {Hirata}  \& {Brinkmann}}{{Mandelbaum} et~al.}{2006}]{mandelbaum2006}
{Mandelbaum} R.,  {Seljak} U.,  {Cool} R.~J.,  {Blanton} M.,  {Hirata} C.~M.,
  {Brinkmann} J.,  2006, \mn@doi [\mnras] {10.1111/j.1365-2966.2006.10906.x},
  \href {https://ui.adsabs.harvard.edu/#abs/2006MNRAS.372..758M} {372, 758}

\bibitem[\protect\citeauthoryear{{Mantz} et~al.,}{{Mantz}
  et~al.}{2016a}]{mant2016}
{Mantz} A.~B.,  et~al., 2016a, \mn@doi [\mnras] {10.1093/mnras/stw2250}, \href
  {https://ui.adsabs.harvard.edu/abs/2016MNRAS.463.3582M} {463, 3582}

\bibitem[\protect\citeauthoryear{{Mantz} et~al.,}{{Mantz}
  et~al.}{2016b}]{mantz2016}
{Mantz} A.~B.,  et~al., 2016b, \mn@doi [Monthly Notices of the Royal
  Astronomical Society] {10.1093/mnras/stw2250}, \href
  {https://ui.adsabs.harvard.edu/abs/2016MNRAS.463.3582M} {463, 3582}

\bibitem[\protect\citeauthoryear{{McClintock} et~al.,}{{McClintock}
  et~al.}{2019a}]{mcclintock2019}
{McClintock} T.,  et~al., 2019a, \mn@doi [\mnras] {10.1093/mnras/sty2711},
  \href {https://ui.adsabs.harvard.edu/\#abs/2019MNRAS.482.1352M} {482, 1352}

\bibitem[\protect\citeauthoryear{{McClintock} et~al.,}{{McClintock}
  et~al.}{2019b}]{mcclintock2019apj}
{McClintock} T.,  et~al., 2019b, \mn@doi [\apj] {10.3847/1538-4357/aaf568},
  \href {https://ui.adsabs.harvard.edu/abs/2019ApJ...872...53M} {872, 53}

\bibitem[\protect\citeauthoryear{{Melchior} et~al.,}{{Melchior}
  et~al.}{2017}]{melchior2017}
{Melchior} P.,  et~al., 2017, \mn@doi [\mnras] {10.1093/mnras/stx1053}, \href
  {https://ui.adsabs.harvard.edu/\#abs/2017MNRAS.469.4899M} {469, 4899}

\bibitem[\protect\citeauthoryear{{Miller} et~al.,}{{Miller}
  et~al.}{2013}]{miller2013}
{Miller} L.,  et~al., 2013, \mn@doi [\mnras] {10.1093/mnras/sts454}, \href
  {https://ui.adsabs.harvard.edu/\#abs/2013MNRAS.429.2858M} {429, 2858}

\bibitem[\protect\citeauthoryear{{Munari}, {Biviano}, {Borgani}, {Murante}  \&
  {Fabjan}}{{Munari} et~al.}{2013}]{munari2013}
{Munari} E.,  {Biviano} A.,  {Borgani} S.,  {Murante} G.,   {Fabjan} D.,  2013,
  \mn@doi [\mnras] {10.1093/mnras/stt049}, \href
  {https://ui.adsabs.harvard.edu/abs/2013MNRAS.430.2638M} {430, 2638}

\bibitem[\protect\citeauthoryear{{Murata}, {Nishimichi}, {Takada}, {Miyatake},
  {Shirasaki}, {More}, {Takahashi}  \& {Osato}}{{Murata}
  et~al.}{2018}]{murata2018}
{Murata} R.,  {Nishimichi} T.,  {Takada} M.,  {Miyatake} H.,  {Shirasaki} M.,
  {More} S.,  {Takahashi} R.,   {Osato} K.,  2018, \mn@doi [The Astrophysical
  Journal] {10.3847/1538-4357/aaaab8}, \href
  {https://ui.adsabs.harvard.edu/abs/2018ApJ...854..120M} {854, 120}

\bibitem[\protect\citeauthoryear{{Murata} et~al.,}{{Murata}
  et~al.}{2019}]{murata2019}
{Murata} R.,  et~al., 2019, arXiv e-prints, \href
  {https://ui.adsabs.harvard.edu/abs/2019arXiv190407524M} {p. arXiv:1904.07524}

\bibitem[\protect\citeauthoryear{{Navarro}, {Frenk}  \& {White}}{{Navarro}
  et~al.}{1996}]{navarro1996}
{Navarro} J.~F.,  {Frenk} C.~S.,   {White} S.~D.~M.,  1996, \mn@doi [\apj]
  {10.1086/177173}, \href {http://adsabs.harvard.edu/abs/1996ApJ...462..563N}
  {462, 563}

\bibitem[\protect\citeauthoryear{{Navarro}, {Frenk}  \& {White}}{{Navarro}
  et~al.}{1997}]{nfw1997}
{Navarro} J.~F.,  {Frenk} C.~S.,   {White} S.~D.~M.,  1997, \mn@doi [\apj]
  {10.1086/304888}, \href {http://adsabs.harvard.edu/abs/1997ApJ...490..493N}
  {490, 493}

\bibitem[\protect\citeauthoryear{{Newman}, {Treu}, {Ellis}, {Sand}, {Nipoti},
  {Richard}  \& {Jullo}}{{Newman} et~al.}{2013}]{newman2013}
{Newman} A.~B.,  {Treu} T.,  {Ellis} R.~S.,  {Sand} D.~J.,  {Nipoti} C.,
  {Richard} J.,   {Jullo} E.,  2013, \mn@doi [\apj]
  {10.1088/0004-637X/765/1/24}, \href
  {http://adsabs.harvard.edu/abs/2013ApJ...765...24N} {765, 24}

\bibitem[\protect\citeauthoryear{{Noh} \& {Cohn}}{{Noh} \&
  {Cohn}}{2012}]{noh2012}
{Noh} Y.,  {Cohn} J.~D.,  2012, \mn@doi [\mnras]
  {10.1111/j.1365-2966.2012.21810.x}, \href
  {https://ui.adsabs.harvard.edu/abs/2012MNRAS.426.1829N} {426, 1829}

\bibitem[\protect\citeauthoryear{{Okabe}, {Takada}, {Umetsu}, {Futamase}  \&
  {Smith}}{{Okabe} et~al.}{2010}]{okabe2010}
{Okabe} N.,  {Takada} M.,  {Umetsu} K.,  {Futamase} T.,   {Smith} G.~P.,  2010,
  \mn@doi [\pasj] {10.1093/pasj/62.3.811}, \href
  {http://adsabs.harvard.edu/abs/2010PASJ...62..811O} {62, 811}

\bibitem[\protect\citeauthoryear{{Okabe}, {Futamase}, {Kajisawa}  \&
  {Kuroshima}}{{Okabe} et~al.}{2014}]{okabe2014}
{Okabe} N.,  {Futamase} T.,  {Kajisawa} M.,   {Kuroshima} R.,  2014, \mn@doi
  [\apj] {10.1088/0004-637X/784/2/90}, \href
  {https://ui.adsabs.harvard.edu/\#abs/2014ApJ...784...90O} {784, 90}

\bibitem[\protect\citeauthoryear{{Pizzuti}, {Saltas}, {Casas}, {Amendola}  \&
  {Biviano}}{{Pizzuti} et~al.}{2019}]{pizzuti2019}
{Pizzuti} L.,  {Saltas} I.~D.,  {Casas} S.,  {Amendola} L.,   {Biviano} A.,
  2019, arXiv e-prints, \href
  {https://ui.adsabs.harvard.edu/\#abs/2019arXiv190101961P} {p.
  arXiv:1901.01961}

\bibitem[\protect\citeauthoryear{{Planck Collaboration} et~al.,}{{Planck
  Collaboration} et~al.}{2016a}]{planck2016}
{Planck Collaboration} et~al., 2016a, \mn@doi [\aap]
  {10.1051/0004-6361/201525830}, \href
  {https://ui.adsabs.harvard.edu/#abs/2016A&A...594A..13P} {594, A13}

\bibitem[\protect\citeauthoryear{{Planck Collaboration} et~al.,}{{Planck
  Collaboration} et~al.}{2016b}]{planck2016SZE}
{Planck Collaboration} et~al., 2016b, \mn@doi [\aap]
  {10.1051/0004-6361/201525833}, \href
  {https://ui.adsabs.harvard.edu/\#abs/2016A&A...594A..24P} {594, A24}

\bibitem[\protect\citeauthoryear{{Press} \& {Schechter}}{{Press} \&
  {Schechter}}{1974}]{press1974}
{Press} W.~H.,  {Schechter} P.,  1974, \mn@doi [\apj] {10.1086/152650}, \href
  {https://ui.adsabs.harvard.edu/abs/1974ApJ...187..425P} {187, 425}

\bibitem[\protect\citeauthoryear{{Raghunathan} et~al.,}{{Raghunathan}
  et~al.}{2019}]{raghunathan2019}
{Raghunathan} S.,  et~al., 2019, \mn@doi [The Astrophysical Journal]
  {10.3847/1538-4357/ab01ca}, \href
  {https://ui.adsabs.harvard.edu/abs/2019ApJ...872..170R} {872, 170}

\bibitem[\protect\citeauthoryear{{Rosati}, {Borgani}  \& {Norman}}{{Rosati}
  et~al.}{2002}]{rosati2002}
{Rosati} P.,  {Borgani} S.,   {Norman} C.,  2002, \mn@doi [\araa]
  {10.1146/annurev.astro.40.120401.150547}, \href
  {http://adsabs.harvard.edu/abs/2002ARA%26A..40..539R} {40, 539}

\bibitem[\protect\citeauthoryear{{Rozo} \& {Rykoff}}{{Rozo} \&
  {Rykoff}}{2014}]{rozo2014}
{Rozo} E.,  {Rykoff} E.~S.,  2014, \mn@doi [The Astrophysical Journal]
  {10.1088/0004-637X/783/2/80}, \href
  {https://ui.adsabs.harvard.edu/abs/2014ApJ...783...80R} {783, 80}

\bibitem[\protect\citeauthoryear{{Rozo} et~al.,}{{Rozo}
  et~al.}{2009}]{rozo2009}
{Rozo} E.,  et~al., 2009, \mn@doi [\apj] {10.1088/0004-637X/699/1/768}, \href
  {https://ui.adsabs.harvard.edu/\#abs/2009ApJ...699..768R} {699, 768}

\bibitem[\protect\citeauthoryear{{Rykoff} et~al.,}{{Rykoff}
  et~al.}{2014}]{rykoff2014}
{Rykoff} E.~S.,  et~al., 2014, \mn@doi [\apj] {10.1088/0004-637X/785/2/104},
  \href {https://ui.adsabs.harvard.edu/#abs/2014ApJ...785..104R} {785, 104}

\bibitem[\protect\citeauthoryear{{Sarazin}}{{Sarazin}}{1986}]{sarazin1986}
{Sarazin} C.~L.,  1986, \mn@doi [Reviews of Modern Physics]
  {10.1103/RevModPhys.58.1}, \href
  {http://adsabs.harvard.edu/abs/1986RvMP...58....1S} {58, 1}

\bibitem[\protect\citeauthoryear{{Saro} et~al.,}{{Saro}
  et~al.}{2015}]{saro2015}
{Saro} A.,  et~al., 2015, \mn@doi [\mnras] {10.1093/mnras/stv2141}, \href
  {https://ui.adsabs.harvard.edu/\#abs/2015MNRAS.454.2305S} {454, 2305}

\bibitem[\protect\citeauthoryear{{Saro} et~al.,}{{Saro}
  et~al.}{2017}]{saro2017}
{Saro} A.,  et~al., 2017, \mn@doi [\mnras] {10.1093/mnras/stx594}, \href
  {https://ui.adsabs.harvard.edu/\#abs/2017MNRAS.468.3347S} {468, 3347}

\bibitem[\protect\citeauthoryear{{Shan} et~al.,}{{Shan}
  et~al.}{2014}]{huanyuan2014}
{Shan} H.~Y.,  et~al., 2014, \mn@doi [\mnras] {10.1093/mnras/stu1040}, \href
  {https://ui.adsabs.harvard.edu/\#abs/2014MNRAS.442.2534S} {442, 2534}

\bibitem[\protect\citeauthoryear{{Shirasaki}, {Takada}, {Miyatake},
  {Takahashi}, {Hamana}, {Nishimichi}  \& {Murata}}{{Shirasaki}
  et~al.}{2017}]{shirasaki2017}
{Shirasaki} M.,  {Takada} M.,  {Miyatake} H.,  {Takahashi} R.,  {Hamana} T.,
  {Nishimichi} T.,   {Murata} R.,  2017, \mn@doi [\mnras]
  {10.1093/mnras/stx1477}, \href
  {https://ui.adsabs.harvard.edu/\#abs/2017MNRAS.470.3476S} {470, 3476}

\bibitem[\protect\citeauthoryear{{Simet}, {McClintock}, {Mandelbaum}, {Rozo},
  {Rykoff}, {Sheldon}  \& {Wechsler}}{{Simet} et~al.}{2017}]{simet2017}
{Simet} M.,  {McClintock} T.,  {Mandelbaum} R.,  {Rozo} E.,  {Rykoff} E.,
  {Sheldon} E.,   {Wechsler} R.~H.,  2017, \mn@doi [\mnras]
  {10.1093/mnras/stw3250}, \href
  {https://ui.adsabs.harvard.edu/\#abs/2017MNRAS.466.3103S} {466, 3103}

\bibitem[\protect\citeauthoryear{{Smith} et~al.,}{{Smith}
  et~al.}{2016}]{smith2016}
{Smith} G.~P.,  et~al., 2016, \mn@doi [\mnras] {10.1093/mnrasl/slv175}, \href
  {https://ui.adsabs.harvard.edu/\#abs/2016MNRAS.456L..74S} {456, L74}

\bibitem[\protect\citeauthoryear{{Takahashi}, {Sato}, {Nishimichi}, {Taruya}
  \& {Oguri}}{{Takahashi} et~al.}{2012}]{takahashi2012}
{Takahashi} R.,  {Sato} M.,  {Nishimichi} T.,  {Taruya} A.,   {Oguri} M.,
  2012, \mn@doi [\apj] {10.1088/0004-637X/761/2/152}, \href
  {https://ui.adsabs.harvard.edu/#abs/2012ApJ...761..152T} {761, 152}

\bibitem[\protect\citeauthoryear{{Tinker}, {Weinberg}, {Zheng}  \&
  {Zehavi}}{{Tinker} et~al.}{2005}]{tinker2005}
{Tinker} J.~L.,  {Weinberg} D.~H.,  {Zheng} Z.,   {Zehavi} I.,  2005, \mn@doi
  [\apj] {10.1086/432084}, \href
  {https://ui.adsabs.harvard.edu/\#abs/2005ApJ...631...41T} {631, 41}

\bibitem[\protect\citeauthoryear{{Tinker}, {Kravtsov}, {Klypin}, {Abazajian},
  {Warren}, {Yepes}, {Gottl{\"o}ber}  \& {Holz}}{{Tinker}
  et~al.}{2008}]{tinker2008}
{Tinker} J.,  {Kravtsov} A.~V.,  {Klypin} A.,  {Abazajian} K.,  {Warren} M.,
  {Yepes} G.,  {Gottl{\"o}ber} S.,   {Holz} D.~E.,  2008, \mn@doi [\apj]
  {10.1086/591439}, \href
  {https://ui.adsabs.harvard.edu/abs/2008ApJ...688..709T} {688, 709}

\bibitem[\protect\citeauthoryear{{Tinker}, {Robertson}, {Kravtsov}, {Klypin},
  {Warren}, {Yepes}  \& {Gottl{\"o}ber}}{{Tinker} et~al.}{2010}]{tinker2010}
{Tinker} J.~L.,  {Robertson} B.~E.,  {Kravtsov} A.~V.,  {Klypin} A.,  {Warren}
  M.~S.,  {Yepes} G.,   {Gottl{\"o}ber} S.,  2010, \mn@doi [\apj]
  {10.1088/0004-637X/724/2/878}, \href
  {https://ui.adsabs.harvard.edu/\#abs/2010ApJ...724..878T} {724, 878}

\bibitem[\protect\citeauthoryear{{Voges} et~al.,}{{Voges}
  et~al.}{1999}]{voges1999}
{Voges} W.,  et~al., 1999, \aap, \href
  {http://adsabs.harvard.edu/abs/1999A%26A...349..389V} {349, 389}

\bibitem[\protect\citeauthoryear{{Warren}, {Abazajian}, {Holz}  \&
  {Teodoro}}{{Warren} et~al.}{2006}]{warren2006}
{Warren} M.~S.,  {Abazajian} K.,  {Holz} D.~E.,   {Teodoro} L.,  2006, \mn@doi
  [The Astrophysical Journal] {10.1086/504962}, \href
  {https://ui.adsabs.harvard.edu/abs/2006ApJ...646..881W} {646, 881}

\bibitem[\protect\citeauthoryear{{Yang}, {Mo}, {van den Bosch}, {Jing},
  {Weinmann}  \& {Meneghetti}}{{Yang} et~al.}{2006}]{yang2006}
{Yang} X.,  {Mo} H.~J.,  {van den Bosch} F.~C.,  {Jing} Y.~P.,  {Weinmann}
  S.~M.,   {Meneghetti} M.,  2006, \mn@doi [\mnras]
  {10.1111/j.1365-2966.2006.11091.x}, \href
  {http://adsabs.harvard.edu/abs/2006MNRAS.373.1159Y} {373, 1159}

\bibitem[\protect\citeauthoryear{{Zhang} et~al.,}{{Zhang}
  et~al.}{2019}]{zhang2019}
{Zhang} Y.,  et~al., 2019, \mn@doi [\mnras] {10.1093/mnras/stz1361}, \href
  {https://ui.adsabs.harvard.edu/abs/2019MNRAS.487.2578Z} {487, 2578}

\bibitem[\protect\citeauthoryear{{Zu} \& {Mandelbaum}}{{Zu} \&
  {Mandelbaum}}{2015}]{zu2015}
{Zu} Y.,  {Mandelbaum} R.,  2015, \mn@doi [Monthly Notices of the Royal
  Astronomical Society] {10.1093/mnras/stv2062}, \href
  {https://ui.adsabs.harvard.edu/abs/2015MNRAS.454.1161Z} {454, 1161}

\makeatother
\end{thebibliography}

% Don't change these lines
\bsp	% typesetting comment
\label{lastpage}
\end{document}